%% file: draft.tex
\documentclass[a4paper, amsfonts, amssymb, amsmath, reprint, showkeys, nofootinbib, twoside, floatfix, pre,superscriptaddress]{revtex4-2}

\usepackage[utf8]{inputenc}
\usepackage[left=23mm,right=13mm,top=35mm,columnsep=15pt]{geometry} 
\usepackage{amsmath}
\usepackage{amssymb}
\usepackage{esint}
\usepackage[unicode=true, pdfusetitle, bookmarks=true, bookmarksnumbered=false, bookmarksopen=false, breaklinks=false, pdfborder={0 0 1}, backref=false, colorlinks=false]{hyperref}
\usepackage{graphicx}
\graphicspath{ {./images/} }

\makeatletter
\pdfoutput=1
\usepackage{makeidx}\usepackage{amsfonts}
\newcommand{\gftwo}{\text{GF}(2)}

\newcommand{\us}{\underline{\sigma}}
\newcommand{\dd}{{\rm d}}
\newcommand{\hm}{\hat{m}}
\newcommand{\lm}{\bar{m}}
\newcommand{\lhm}{\bar{\hat{m}}}
\newcommand{\di}{\partial i}
\newcommand{\da}{\partial a}
\newcommand{\dima}{\partial i \setminus a}
\newcommand{\dami}{\partial a \setminus i}
\newcommand{\tl}{\tilde{\lambda}}
\newcommand{\tp}{\tilde{p}}
\newcommand{\cP}{\mathcal{P}}
\newcommand{\cQ}{\mathcal{Q}}
\newcommand{\cG}{\mathcal{G}}

\DeclareMathOperator*{\argmin}{arg\,min}
\DeclareMathOperator*{\argmax}{arg\,max}

\makeatother

\begin{document}
\title{The closest vector problem and the zero-temperature p-spin landscape for lossy compression}

% \affiliation{DISAT, Politecnico di Torino, Corso Duca Degli Abruzzi 24, I-10129 Torino, Italy}
% {IIGM}
% \affiliation{Dipartimento di Fisica, Sapienza Universit\`a di Roma, P.le A. Moro 5, Rome 00185 Italy}{SAPIENZA}
% \affiliation{CNR--Nanotec, Rome unit, P.le A. Moro 5, Rome 00185 Italy}{CNR}
% \affiliation{INFN--Sezione di Roma1, P.le A. Moro 5, Rome 00185 Italy}{INFN}

\author{Alfredo Braunstein}
\affiliation{DISAT, Politecnico di Torino, Corso Duca Degli Abruzzi 24, I-10129 Torino, Italy}

\author{Louise Budzynski}
\affiliation{DISAT, Politecnico di Torino, Corso Duca Degli Abruzzi 24, I-10129 Torino, Italy}
\affiliation{Italian Institute for Genomic Medicine, IRCCS Candiolo, SP-142, I-10060, Candiolo (TO), Italy}

\author{Stefano Crotti}
\affiliation{DISAT, Politecnico di Torino, Corso Duca Degli Abruzzi 24, I-10129 Torino, Italy}

\author{Federico Ricci-Tersenghi}
\affiliation{Dipartimento di Fisica, Sapienza Universit\`a di Roma, P.le A. Moro 5, Rome 00185 Italy}
\affiliation{CNR--Nanotec, Rome unit, P.le A. Moro 5, Rome 00185 Italy}
\affiliation{INFN--Sezione di Roma1, P.le A. Moro 5, Rome 00185 Italy}

\begin{abstract}
\input{sections/abstract}
\end{abstract}
\maketitle
\tableofcontents
\input{sections/introduction}
\input{sections/definitions}
\input{sections/deg2_results}

\input{sections/deg3_results}
\input{sections/cavity_method_results}
\input{sections/conclusions}
\appendix
\input{sections/gfq}

\input{sections/cavity_method_equations}
\input{sections/sparsebasis}

\input{sections/ldpc}

\bibliography{draft}

\end{document}

%% file: sections/abstract.tex
We consider a high-dimensional random constrained optimization problem in which a set of binary variables is subjected to a linear system of equations.
The cost function is a simple linear cost, measuring the Hamming distance with respect to a reference configuration.
Despite its apparent simplicity, this problem exhibits a rich phenomenology. 
We show that different situations arise depending on the random ensemble of linear systems.
When each variable is involved in at most two linear constraints, we show that the problem can be partially solved analytically, in particular we show that upon convergence, the zero-temperature limit of the cavity equations returns the optimal solution.
We then study the geometrical properties of more general random ensembles. 
In particular we observe a range in the density of constraints at which the systems enters a glassy phase where the cost function has many minima.
Interestingly, the algorithmic performances are only sensitive to another phase transition affecting the structure of configurations allowed by the linear constraints.
We also extend our results to variables belonging to $\text{GF}(q)$, the Galois Field of order $q$.
We show that increasing the value of $q$ allows to achieve a better optimum, which is confirmed by the Replica Symmetric cavity method predictions.

%% file: sections/introduction.tex
\section{Introduction}
The Closest Vector Problem (CVP) is a constrained optimization problem in which the objective is to find the vector in a high-dimensional lattice that is closest to a given reference vector (in principle, external to the lattice). 
In this work we will study lattices defined by linear subspaces of $\text{GF}(q)^n$, the $n$-dimensional vector space over the Galois Field $\text{GF}(q)$. 
In particular for the case $q=2$ (which we will mainly study), the lattice is defined as the solution set of a XORSAT instance, and the CVP can also be recast in the problem of finding the ground state of an Ising spin glass model with external fields.

CVP is a fundamental problem in theoretical computer science: some versions of CVP were among the first ones where an equivalence between worst case and average case complexity has been shown. Such an equivalence has been exploited to propose a robust cryptosystem \cite{ajtai1997public} .
Nonetheless the CVP has been proven to be computationally NP-hard \cite{ajtai1998shortest} and hard to approximate, even allowing a potentially slow pre-processing step \cite{cvphardness, feige2004inapproximability}.

The main motivation for studying this model is that it is an ideal framework to understand, via statistical mechanics computations, the mechanisms underlying the hardness in approximating optimal solutions.
As we will see, it shows a wide range of non-trivial properties in the geometry of the solution landscape, while (partially) conserving some analytic feasibility coming from the fact that its configurations are solutions of a linear system in $\text{GF}(q)$.

Surprisingly, we will show that approximating optimal solutions is hard even in a region of parameters where the space of solutions is ``well connected''. In this region the model corresponding to the uniform measure over solutions looks nice enough (e.g. it is in a paramagnetic phase, correlations decay fast enough, a replica-symmetric solution well describes the Gibbs measure).
Nonetheless when we add the external field to search for the closest solution to a reference vector, the scenario changes dramatically: ergodicity breaking phase transitions take place and the problem becomes very hard.

Given that the search space is always the same (solutions to a XORSAT constraint satisfaction problem, CSP) the addition of the external field can be seen as a reweighting of the CSP solution space.
It is well known that the reweighting of the solution space can induce ergodicity breaking phase transitions \cite{ricci2009cavity} and change the location of the phase transitions \cite{budzynski2019biased,cavaliere2021optimization}.
In the present model we are going to show how important the effects of the reweighting can be and how they can affect algorithms searching for optimal solutions, which are relevant is several common applications.

\subsection{Compression}
\label{subsec:compression}
The CVP has a straightforward application to the lossy compression of a symmetric binary source (source coding). 
In this context, the compression task is to take an input $\underline{y}\in \text{GF}(2)^n$ and to reduce it into a compressed version $\underline{c}\in \text{GF}(2)^k$ with $k<n$. The decompression task transforms $\underline{c}$ into $\underline{\hat{x}}\in \text{GF}(2)^n$. The distortion is defined as the Hamming distance $d_H(\underline{y},\underline{\hat{x}})$, i.e. the number of differing components between the two vectors. A good compression scheme is designed to result in the smallest possible distortions, and the performance of the scheme can be measured in terms of average distortion on random binary inputs.

As an example, one trivial compression scheme consists in truncating the input to the first $k$ components (compression) and reconstructing randomly the last $n-k$ components (decompression). 
Of course, it is possible to do much better. In fact, the best possible performance for a given input distribution has been characterized by Shannon \cite{shannon1959coding, shannon1948mathematical,shannon1957certain} thanks to a duality with the channel coding problem. The smallest achievable average distortion for a given pair $n,k$ on a binary symmetric source is given by the equation 
\begin{align}
	\label{eq:rate-dist-bound}
	D=H^{-1}\left(1-R\right)
\end{align} 
where $R=\frac{k}{n}$ is called the compression rate, and where $$H(p)=-p\log_2(p)-(1-p)\log_2(1-p)$$ is the binary entropy function and $H^{-1}$ is its inverse. Interestingly, a theoretical asymptotically optimal (but computationally inefficient) scheme is formed by random codes. Random codes are constructed as follows: choose $2^k$ random vectors $\underline{v}_1,\dots,\underline{v}_{2^k}$ in $GF(2)^n$. The compression scheme consists in finding the vector $\underline{v}_i$ that is closest to $\underline{y}$. 
The binary representation $\underline{c}$ of $i$ will be the compressed vector. When $n,k\to\infty$ with fixed compression rate $R=\frac{k}n$, the average distortion $\left<d_h(\underline{v}_i,\underline{y})\right>$ falls on the optimal line \cite{shannon1959coding}. As it happens with the dual channel coding problem, a computationally more efficient alternative can be constructed by replacing random vectors by solutions of a random linear system (in a discrete space), and this is the approach that has been taken in \cite{braunstein_efficient_2011} and that we will take here.
In particular for $q=2$, random codes correspond to the solution set of instances of a random XORSAT ensemble.
%% maybe add here plot of the two curves?

\subsection{Relation with previous works}
Let $H$ be an $m\times n$ matrix with binary entries, and let $\underline{b}\in\{0,1\}^m$ be a $m$-component vector.
An instance of the XORSAT problem is given by a pair $(H,\underline{b})$, the solution set of this instance being the set of vectors $\underline{x}\in\{0,1\}^n$ satisfying the equation $H\underline{x}=\underline{b}$ modulo 2.
The random XORSAT ensemble is defined by taking $\underline{b}$ uniformly at random in $\{0,1\}^m$, and by taking $H$ from some random matrix ensemble.
In this paper, we will study the ensemble in which $H$ is sampled uniformly over the set of matrices having a prescribed distribution in the number of non-zero entries per rows and per columns.
The random XORSAT problem has been studied extensively in the past \cite{mezard2003two, dubois}. 

Note that the lossy compression problem of a random uniform binary source is formally identically to the decoding problem in a binary symmetric channel (see e.g. \cite{mackay2003information}). Although decoding using low-density parity-check codes (LDPC) has been studied with the cavity method in the past \cite{nakamura_statistical_2001, FranzLeone_2002}, the two problems differ on the ensembles of source vectors. Indeed, in the channel coding problem the source vector is formed by the perturbation of a codeword instead of being a random uniform vector. In a sense, the compression problem then corresponds to a channel coding problem in a regime with non-vanishing probability of decoding errors instead of a vanishing one. This difference makes the compression problem fundamentally harder (see Appendix \ref{sect:ldpc}), and the results not directly comparable. Note that the ensembles of ``good'' codes for both problems also fundamentally differ: for the channel coding problem, having codewords that are ``close together'' renders the code useless (as the error rate will be non vanishing), while for the compression problem it is essentially harmless (and may be indeed convenient from a computational point of view).
%% cite LDPC works

A striking feature of random CSPs is the appearance of phase transitions, or threshold phenomena in the thermodynamic limit, when the number of variables $n$ and the number of linear constraints $m$ go to infinity at a fixed value of the ratio $\alpha=m/n$, the density of constraints per variable.
For instance the satisfiability threshold $\alpha_{\rm sat}$ separates a satisfiable regime $\alpha<\alpha_{\rm sat}$ where random XORSAT instances do admit solutions from an unsatisfiable phase $\alpha>\alpha_{\rm sat}$ where no solution typically exists.
Another transition occurs in the satisfiable phase at $\alpha_d<\alpha_{\rm sat}$, called the clustering transition.
Below $\alpha_d$ the solution set of typical instances is rather well connected, any solution can be reached by any other through a path of nearby solutions.
Above $\alpha_d$ the solution set splits into a exponential number of distinct groups of solutions, called clusters, which are internally well connected, but well separated one from the other.
This transition also manifests itself with the appearance of a specific type of correlations between variables, known as point-to-set correlations, under the uniform probability measure over the set of solutions. These correlations forbid the rapid equilibration of stochastic processes that respect the detailed balance condition~\cite{MontanariSemerjian06b}, which justifies the alternative `dynamic' name of the clustering transition.

In \cite{braunstein_efficient_2011}, a Belief Propagation (BP) scheme has been employed on an ensemble of linear codes called cycle codes. These cycle codes correspond to systems of linear equations (in $\text{GF}(q)$) in which each variable participates in at most two equations. It has been observed that the performance of BP improves by adding some leaves (variables of degree one) to the linear system, but not too many of them. 
In this work, we compute the analytic achievable performance of such codes through the cavity method, and provide a rigorous proof of the exactness of the zero-temperature version of the cavity equations on cycle codes.
We then extend our results to codes with higher degrees, and show that this allows to improve the performance of lossy compression. 
We study the clustering transition for the constrained optimization problem CVP, and show that its clustering threshold $\alpha_d^{\rm CVP}$ arises for density of constraints smaller than the clustering threshold $\alpha_d^\oplus$ associated to the random XORSAT problem defining the set of constraints.
We also study the performances of message-passing algorithms designed to solve this constrained optimization problem.
Interestingly, we observe that these algorithms are not affected by the clustering transition associated to the constrained optimization problem (occurring at $\alpha_d^{\rm CVP}$), but instead are only affected by the XORSAT clustering transition occurring at $\alpha_d^\oplus$.
%In addition, we provide an explanation on the fact that a small number of leaves worsens the code but is still (algorithmically) beneficial.   

\subsection{Relation between sparse basis and clustering}
\label{subsec:clustering_and_sparse_basis}
Given a basis of the solution space, we define the weight of the basis as the maximum Hamming weight of its elements.
We can now establish a fundamental relation between a geometrical property of the solution space and the weight of the lightest basis: for a linear system in $\text{GF}(2)$, having a basis with weight $\leq k$ is equivalent to having all solutions of the system connected between each other by `paths' of solutions with jumps between consecutive ones of hamming distance $\leq k$.
This can be proven easily. Suppose indeed that we have a basis $B=\{\underline{b}_1,\dots \underline{b}_d\}$ in which each element has Hamming weight $\leq k$. Take any two solution vectors $\underline{v},\underline{v}'$, and write in the basis $\underline{v}'-\underline{v}=\sum_{i=1}^d c_i \underline{b}_i$ where $c_i\in GF(2)$. Let $\{i_1,i_2,...,i_r\}=\{i:c_i=1\}$. Construct the sequence $\underline{v}_0 = \underline{v}$, $\underline{v}_j = \underline{v}_{j-1} + \underline{b}_{i_{j}}$ for $1\leq j \leq r$. The difference between two points in the sequence is a basis element, so has weight $\leq k$, and the sequence forms a path from $\underline{v}$ to $\underline{v}'$. Conversely, suppose that all solutions can be connected to the vector $\underline{0}$ with paths of jumps $\underline{\Delta}_j = \underline{v}_j - \underline{v}_{j-1}$ with hamming distance $\leq k$. The set of all those vectors $\underline{\Delta}_j$'s thus spans the full set of solutions. It is possible then to extract a solution basis from such a set, and all elements in the basis will have Hamming weight $\leq k$. In summary, the weight of the sparsest basis determines the largest separation between ``clusters'' or groups of solutions. If the weight of the sparsest basis is $k$, then there will be at least a subset of solutions which is separated by hamming distance $k$ from the rest.

\subsection{Organization}
The paper is organized as follows. 
In section~\ref{sec:definitions} we define more precisely the constrained optimization problem under study and the statistical physics model associated to it. We present the equations that describe its behavior in the framework of the cavity method from statistical mechanics, and the algorithms that we used to solve its instances. 
In section~\ref{sec:deg2_results} we study the case of cycle codes, i.e. when the binary variables are involved in at most $2$ linear constraints. 
In this particular case we prove that when it converges, Max-Sum algorithm finds the optimal solution. 
Moreover, we design an exact greedy algorithm, that we call GO for `greedy optimal' that is guaranteed to converge to the optimal solution. 
In section~\ref{sec:deg3_results}, we study a random ensemble in which the binary variables are involved in more than $2$ linear constraints, and we show that in this ensemble the average minimal distance is smaller than the one for cycle codes, thus providing a better code for lossy compression.
We show that CVP undergoes a clustering transition before the clustering transition associated to the XORSAT problem representing the constraints.
We study the behavior of three algorithms: Belief-Propagation with decimation, Max-Sum with reinforcement, and Survey-Propagation with decimation, and show that their performances are only affected by the clustering transition associated to the XORSAT problem. 
In section~\ref{sec:more_detailed_v}, we provide a more detailed picture of the phase diagram obtained in section~\ref{sec:deg3_results}.
In particular we perform a finite temperature study to relate the two clustering transitions occurring at $\alpha_d^{\rm CVP}$ and $\alpha_d^\oplus$ that correspond respectively to the cases of zero and infinite temperature. 
We also argue in favor of a full RSB transition occurring at higher densities of constraints, that prevented us to obtain a reliable prediction of the minimal distortion in this regime.
In appendix~\ref{sec:gfq} we present the results obtained with variables in $\text{GF}(q)$, the Galois Field of order $q$. We show that cycle codes with higher value of $q$ allows to achieve smaller average minimal distortion, thus providing better codes for lossy compression. This trend is confirmed by the zero-temperature Replica Symmetric prediction. We however argue in favor of a RSB transition as $q$ increases, that could prevent from an efficient compression scheme when $q$ becomes large.
The RS and 1RSB formalism specified for the CVP problem is given in Appendix \ref{sec:cavity_method_equations}.
In appendix~\ref{sect:sparsebasis} we show that for cycle codes with binary variables, it is possible to build a basis whose weight is upper bounded by minimal-size rearrangements computed in \cite{MontanariSemerjian06} by A. Montanari and G. Semerjian.
When this upper-bound remains finite in the thermodynamic limit, this allows us to conclude from the discussion in~\ref{subsec:clustering_and_sparse_basis} that the solution set of XORSAT is well-connected.

%% file: sections/definitions.tex
\section{Definition of the model and statistical mechanics formalism}
\label{sec:definitions}

\subsection{Definition of the model}
The constrained optimization problem can be formulated as follows. Given a reference vector $\underline{y}\in\{0,1\}^n$ and a linear subspace $\mathcal{C}\subset\{0,1\}^n$, find a vector $\underline{\hat{x}}\in\mathcal{C}$ that is the closest to $\underline{y}$:
\begin{equation}\label{eq:min2}
\hat{\underline{x}} = \argmin_{\underline{x}\in \mathcal{C}} d_H(\underline{x},\underline{y})
\end{equation}
where 
\begin{align}
	d_H(\underline{x},\underline{y}) = \frac{1}{n}\sum_{i=1}^{n} x_i \oplus y_i
\end{align} 
is the Hamming distance.
The linear subspace $\mathcal{C}$ is defined as the solution set of an homogeneous XORSAT instance: let $H$ be an $m\times n$ matrix with boolean entries $H_{ia}\in\{0,1\}$, $i\in\{1,\dots, n\}$, $a\in\{1,\dots, m\}$, then:
\begin{align}
	\mathcal{C}=\{\underline x\in\{0,1\}^n: H\underline x=0\} \ .
\end{align}
Here homogeneous means that the r.h.s.\ of the linear system is equal to the null vector $\underline{b}=\underline{0}$.
One can encode the topology of a XORSAT instance into a bipartite graph $\cG=(V,F,E)$. 
The set of variable nodes $V$ represents the binary variables $x_1,\dots,x_n$. The set of factor nodes $F$ represents the $m$ constraints encoded in the $m$ rows of $H$.
An edge $(i,a)\in E$ is drawn if variable $x_i$ is involved in the $a$-th constraint: $H_{ia}=1$.
By means of a simple change of variables,
\begin{equation}
    \begin{aligned}
    \sigma_i &= (-1)^{x_i}\\
    s_i &= (-1)^{y_i}\\
    \end{aligned}
\end{equation}
the problem can be re-written as a statistical physics model. We define the probability law:  
\begin{align}\label{eq:prob}
    \mu(\us) = \frac{1}{Z(\beta)}\left(\prod_{a=1}^m\mathbb{I}\left[\prod_{i\in\da}\sigma_i=1\right]\right)e^{\beta\sum_{i=1}^ns_i\sigma_i}
\end{align}
with $\mathbb{I}[A]$ being the indicator function of the event $A$. The problem (\ref{eq:min2}) is then equivalent to finding the configuration $\us$ maximizing the probability law $\mu(\us)$, or equivalently minimizing the energy function 
\begin{align}
	\label{eq:energy}
    E(\sigma)= -\sum_{i=1}^n\sigma_i s_i
\end{align}
under the set of constraints $\{\prod_{i\in\da}\sigma_i=1\}_{a=1}^m$. Note that $E(\us)$ can be related to the distortion as follows: $E(\us)=2d_H(\underline{x},\underline{y})-1$.

It will be convenient to define a softened version of the probability law $\mu(\us)$, by replacing the hard constraints on the factors $\mathbb{I}\left[\prod_{i\in\da}\sigma_i=1\right]$ by a soft constraint $e^{\beta J\left(\prod_{i\in\da}\sigma_i -1\right)}$, with $J$ a real parameter:
\begin{align}\label{eq:prob_J}
    \mu_J(\us) = \frac{1}{Z(\beta,J)}e^{-\beta E_J(\us)}
\end{align}
where:
\begin{align}\label{eq:MinNrjcutoffJ}
E_J(\us) = -J\sum_{a=1}^m\left(\prod_{i\in\da}\sigma_i -1\right) - \sum_{i=1}^n\sigma_i s_i \ .
\end{align}
The first term is bringing an energetic cost $2J$ to each unsatisfied clause, while the second term is the original energy function, which favors configurations close to the source.
In statistical physics, this model is known as a spin glass model in presence of heterogeneous external fields $s_1,\dots,s_n$.
Sending $J\to\infty$ allows to recover the probability law $\mu(\us)$ defined in (\ref{eq:prob}). 

\subsection{Random ensemble of instances}

We will be interested in the characterization of the `typical' properties of this constrained optimization problem, where typical is defined with respect to a random ensemble of instances, a property being considered typical if it occurs with a probability going to one in the thermodynamic (large size) limit. 
In particular, we will consider random external fields $s_1, \dots, s_n$ in which each external field $s_i$ is i.i.d.\ uniformly in $\{-1,1\}$.
As we have seen in the previous subsection, the set of constraints can be represented by a bipartite graph $\cG=(V,F,E)$.
We will consider random graph ensembles with fixed degree profiles, denoted $\mathbb{G}_n(\Lambda, P)$. 
Let $\Lambda=\{\lambda_1,\dots,\lambda_{d_{\max}}\}$ be the degree profile of the variable nodes, with $d_{\max}$ the maximal degree, and $\lambda_i$ the fraction of variable nodes of degree $i$.
Respectively, let $P=\{p_1,\dots, p_{k_{\max}}\}$ be the degree profile of the factor nodes, with $k_{\max}$ the maximal degree, and $p_i$ the fraction of factor nodes of degree $i$.
The degree profiles are normalized: $\sum_{i=1}^{d_{\max}}\lambda_i=1$ and $\sum_{i=1}^{k_{\max}}p_i=1$, and they satisfy the following relation $$m\sum_{i=1}^{k_{\max}}ip_i=n\sum_{i=1}^{d_{\max}}i\lambda_i =|E| \ .$$
We will be interested in the thermodynamic limit $n, m\to \infty$, with a fixed ratio $\alpha=m/n$, and fixed fractions $\lambda_i,p_i$'s independent of $n$.
The ratio $\alpha$ is called the density of constraints per variable and is related to the degree profiles as follows: $\alpha = \frac{\sum_{i=1}^{d_{\max}}i\lambda_i}{\sum_{i=1}^{k_{\max}}ip_i}$.
In the thermodynamic limit, random graphs extracted from $\mathbb{G}_n(\Lambda, P)$ are locally tree-like: the neighborhood of an uniformly chosen vertex within a finite distance is acyclic, with probability going to $1$.
Note that in the formalism of lossy compression, the compression rate $R=\frac{n-m}{m}$ can be expressed in terms of the degree profiles:
\begin{align}
	R=1-\alpha = 1-\frac{\sum_{i=1}^{d_{\max}}i\lambda_i}{\sum_{i=1}^{k_{\max}}ip_i}
\end{align}
The equivalence between the CVP and the spin glass model with external fields allows us to apply the cavity method. This method has been first developed in the context of statistical physics of disordered systems, and has later on been applied to random Constraint Satisfaction Problems. 
The aim of the cavity method is to characterize the properties of the probability measure (\ref{eq:prob}), for typical random graphs in $\mathbb{G}_n(\Lambda, P)$ and realization of the external fields $s_1,\dots,s_n$, in the thermodynamic limit.
In particular, we will be interested in the zero-temperature limit of the cavity method, at which the probability measure (\ref{eq:prob}) concentrates on the configurations satisfying the constraints and achieving the minimal energy.
A simplified version of the cavity method, especially of the 1RSB formalism, first derived in \cite{MezardParisi03, MezardZecchina02}, can be obtained in this limit.
We will also be interested in the finite-temperature (or finite-$\beta$) version of the cavity method (see \ref{subsec:finite_T_study}). 
We give the details of the cavity method applied to the CVP in the appendix~\ref{sec:cavity_method_equations}.

\subsection{BP equations and Bethe free-energy}

Belief-Propagation (BP) is a method that allows to study the properties of the measure $\mu$ defined in (\ref{eq:prob}) on a single instance, and at finite inverse temperature $\beta$.
When the bipartite graph $\cG$ representing the constraints is a tree, this method is exact, and allows to compute the partition function $Z(\beta)$, as well as the marginal probabilities of any variable $\sigma_i$.
In practice, the BP method is also used as a heuristic on random sparse instances.
For each variable node $i\in V$, we denote by $\di=\{a\in F: (ia)\in E\}$ the set of factor nodes connected to $i$, and similarly for each factor node $a\in F$ the set of variable nodes connected to $a$: $\da=\{i\in V: (ia)\in E\}$. 
We introduce the Belief-Propagation (BP) messages $m_{i\to a}$ and $\hm_{a\to i}$ on each edge $(i,a)\in E$ as the marginal probability laws of $\sigma_i$ in the amputated graph where some interactions are discarded: $m_{i\to a}$ is the marginal of $\sigma_i$ when the hyperedge $a$ is removed, and $\hm_{a\to i}$ is the marginal of $\sigma_i$ when one removes all the hyperedges in $\dima$.
The BP messages obey the following recursive equations:
\begin{align}\label{eq:BP}
\begin{aligned}
    m_{i\to a}(\sigma_i) &= \frac{1}{z_{i\to a}}e^{\beta\sigma_i s_i}\prod_{b\in\dima}\hm_{b\to i}(\sigma_i)\\
    \hm_{a\to i}(\sigma_i) &= \frac{1}{\hat{z}_{a\to i}}\sum_{\us_{\dami}} \mathbb{I}\left[\prod_{i\in\da}\sigma_i=1\right]\prod_{j\in\dami}m_{j\to a}(\sigma_j)
\end{aligned}
\end{align}
where $z_{i\to a}, \hat{z}_{a\to i}$ are normalization factors. One can compute the marginal probability of $\sigma_i$ from the solution of the above set of equations:
\begin{align}
    \mu_i(\sigma_i) = \frac{1}{z_i}e^{\beta s_i\sigma_i}\prod_{a\in\di}\hm_{a\to i}(\sigma_i)
\end{align}

The Free Energy $F = -(1/\beta)\log(Z(\beta))$ can be expressed in terms of BP messages using the Bethe formula:
\begin{eqnarray}
F^{\rm Bethe}(\underline{m}, \underline{\hm}) &=& \sum_{(i,a)\in E} \frac{1}{\beta}\log Z_{ia}(m_{i\to a},\hm_{a\to i}) \nonumber \\
&& -\sum_{a=1}^m \frac{1}{\beta}\log Z_a(\{m_{i\to a}\}_{i\in\da}) \nonumber \\
&& - \sum_{i=1}^n \frac{1}{\beta}\log Z_i(\{\hm_{a\to i}\}_{a\in\di}) 
\label{eq:BetheF}
\end{eqnarray}
where $Z_a,Z_i,Z_{ia}$ are defined as follow:
\begin{align}
\label{eq:BetheF_defterms}
    Z_a &= \sum_{\us_{\da}}\mathbb{I}\left[\prod_{i\in\da}\sigma_i=1\right]\prod_{i\in\da}m_{i\to a}(\sigma_i)\\
    Z_i &= \sum_{\sigma_i}e^{\beta s_i,\sigma_i}\prod_{a\in\di}\hm_{a\to i}(\sigma_i)\\
    Z_{ia} &= \sum_{\sigma_i}m_{i\to a}(\sigma_i)\hm_{a\to i}(\sigma_i)
\end{align}
Finally, one can also compute the average energy in terms of BP beliefs:
\begin{align}
	\langle E(\us)\rangle_{\mu} = -\sum_{i=1}^n\sum_{\sigma_i}s_i\sigma_i\mu_i(\sigma_i)
\end{align}

\subsection{Zero-temperature limit: Max-Sum equations and Bethe energy}
\label{sec:ms}
The Max-Sum (MS) equations can be seen as the zero-temperature limit of the BP equations. 
The goal here is to describe the set of configurations $\us$ which maximize the probability $\mu(\us)$ in (\ref{eq:prob}), i.e. which solves the constrained optimization problem (\ref{eq:min2}).
We define Max-Sum messages as:
\begin{align}
\begin{aligned}
    h_{i\to a} &= \lim_{\beta\to \infty}\frac{1}{2\beta}(\log m_{i\to a}(+) - \log m_{i\to a}(-)) \\
    u_{i\to a} &= \lim_{\beta\to \infty}\frac{1}{2\beta}(\log \hm_{a\to i}(+) - \log \hm_{a\to i}(-))
\end{aligned}
\end{align}
Using this definition and the BP equations we get the following MS equations (associated to the probability law $\mu(\us)$ defined in (\ref{eq:prob}) for hard constraints):
\begin{align}\label{eq:MS}
\begin{aligned}
    h_{i\to a} &= s_i + \sum_{b\in\dima} u_{b\to i} \\
    u_{a\to i} &= {\rm sign}\left(\prod_{j\in\dami} h_{j\to a}\right)\min_{j\in\dami} (|h_{j\to a}|)
\end{aligned}
\end{align}
Once a solution to the MS equations is found, one can compute the Max-Sum belief $h_i$:
\begin{align}
\begin{aligned}
	\label{eq:MS_belief}
    h_i&=\lim_{\beta\to\infty}\frac{1}{2\beta}(\log(b_i(+))-\log(b_i(-)))\\
    &=s_i + \sum_{a\in\di} u_{a\to i}
\end{aligned}    
\end{align}
which corresponds in case of a tree to the difference in energy $\Delta E_i = E_{i}^{\min}(+)-E_{i}^{\min}(-)$, where $E_{i}^{\min}(\sigma)$ is the ground state energy when $\sigma_i$ is fixed to the value $\sigma$.
Since the energy function $E(\us)$ takes only integer values, one can deduce that the Max-Sum messages satisfying equations (\ref{eq:MS}) and Max-Sum beliefs (corresponding to differences in energy) also take integer values. 
Replacing the hard constraints by soft constraints is equivalent to introducing a cut-off on the values of the factor-to-variable $u_{a\to i}\in[-J,J]$. 
The variable-to-factor messages then take values $h\in[-1-J(d_{\max}-1),1+J(d_{\max}+1)]$. 
Note that the MS equation (\ref{eq:MS}) on $u_{a\to i}$ is replaced by 
\begin{align}
	\label{eq:MS_J}
    u_{a\to i} &= {\rm sign}\left(\prod_{j\in\dami} h_{j\to a}\right)\min\left(\min_{j\in\dami} |h_{j\to a}|, J\right)
\end{align}

% One can compute the minimal energy $E^{\min}$ as the large $\beta$ limit of the Free Energy $F(\beta,J)$. The Bethe Energy is the large $\beta$ limit of the Bethe Free Energy expressed in equation (\ref{eq:BetheF}):
One can compute the minimal energy $E^{\min}$ as the large $\beta$ limit of the Free Energy (\ref{eq:BetheF}):
\begin{align}
    E^{\rm Bethe}(\underline{h}, \underline{u}) = &\sum_{a=1}^m E_a(\{h_{i\to a}\}_{i\in\da}) \nonumber \\
    +&\sum_{i=1}^n E_i(\{u_{a\to i}\}_{a\in\di}) \nonumber \\
    +& \sum_{(i,a)\in E} E_{ia}(h_{i\to a},u_{a\to i}) \label{eq:EBethe} \;,
\end{align}
which is exact when the factor graph is a tree. In the above expression $E_i, E_a, E_{ia}$ are defined as follows:
\begin{align}
    E_a(\{h_{i\to a}\}_{i\in\da}) = & 2\min_{i\in\partial a}(|h_{i\to a}|)\Theta\left(\prod_{i\in\partial a}h_{i\to a}\right) \nonumber \\
    E_i(\{u_{a\to i}\}_{a\in\di}) = & -\left|s_i + \sum_{a\in\di} u_{a\to i}\right| + \sum_{a\in\di}|u_{a\to i}| \nonumber \\
    E_{ia}(h_{i\to a},u_{a\to i}) = &- |u_{a\to i} + h_{i\to a}| \nonumber \\
    &+ |u_{a\to i}| + |h_{i\to a}| \label{eq:EBethe_defterms}
\end{align}
With soft constraints, i.e.\ $J$ finite, the factor contribution $E_a(\{h_{i\to a}\}_{i\in\da})$ to the Bethe minimal energy (\ref{eq:EBethe}) is replaced by
\begin{equation}
    E_a(\{h_{i\to a}\}_{i\in\da}) = 2\min \left(J,\min_{i\in\partial a}|h_{i\to a}|\right)
    \Theta\left(\prod_{i\in\partial a}h_{i\to a}\right) \; .\label{eq:EBethe_defterms_J}
\end{equation}

\subsection{Decimation}
\label{subsec:decimation}
% In this subsection we describe the algorithm based on Belief Propagation with decimation that we used to solve single instances of the Closest Vector Problem.
% Decimation is a sequential assignment procedure in which a variable is fixed to a given value at each time step, until all values are fixed.
% Beliefs are estimates of single-site marginals.
The output of the BP algorithm is just an estimate of single-site marginals, and to find a solution to the optimization problem, one needs to convert these marginals into a specific spin configuration. However, note that picking $\underline\sigma ^*$ where $\sigma^*_i=\argmax b_i(\sigma_i)$ does not lead to a good result in general, as it disregards existing correlations between variables (e.g.\ in case of problems with hard constraints this strategy can lead to inconsistencies, since $\underline\sigma^*$ in general does not satisfy the constraints).
To overcome this issue one typically resorts to decimation, i.e.\ a sequential assignment of the variables according to their beliefs. 

In practice, we use the fact that the set of XORSAT constraints is a linear system of equations to improve our algorithm.
We first build a basis for this linear system (e.g.\ by means of Gaussian elimination), thereby identifying a subset of independent variables.
The decimation procedure is then applied only to these independent variables. Once all independent variables are fixed, the remaining variables are determined by the linear constraints, thus ensuring that we obtain a solution to the linear system.
At each time step, the algorithm solves iteratively the Belief Propagation equations (\ref{eq:BP}) and computes the marginal probabilities of each variable.
Then, the algorithm picks the most biased variable, i.e.\ $i^* = \argmax_i(\mu_i(+)-\mu_i(-))$ among the independent variables that are not yet decimated, samples $\sigma_i \in \{-1, 1\}$ according to its marginal $\mu_i$ and switches on a strong external field in the corresponding direction, in such a way that $\sigma_i$ is now fixed in the direction of its belief.
We worked with sufficiently large value of $\beta$ (in practice we used $\beta=3$), such that the measure (\ref{eq:prob}) is reasonably concentrated around configurations achieving the minimal energy $E(\us)$. In the limit of $\beta\to \infty$ (Max-Sum equations), the system should be fully concentrated on the configurations of minimal energy. If the minimum is not unique, then decimation is still needed to break the symmetry between equivalent ground states. Alternatively, one can add a small symmetry-breaking random external field so that the ground state becomes unique. This is the strategy we adopted for Max-Sum.

\subsection{Reinforcement}

An alternative to decimation is reinforcement, which consists in updating the external field on a variable according to its belief, thus guiding the system to full polarization. Reinforcement is also sometimes called soft decimation, as it sets at each iteration a soft external field of all variables instead of a strong field on only one variable.

The reinforcement procedure can also be employed to help convergence of MS equations, the small external fields accumulate during time (before convergence) to drive the system to a model with strong external fields for which convergence is easier to achieve.

% At each time step, a solution to the Max-Sum equations (\ref{eq:MS}) is found iteratively, and the external field of riable is updated:
At each iteration $t$ of the Max-Sum algorithm \eqref{eq:MS},  the external field on each variable is updated according to its belief
\begin{align}
	s_i^{(t+1)} = s_i^{(t)} + \gamma(t)h_i^{(t)} 
\end{align}
with $h_i^{(t)}$ the Max-Sum belief (\ref{eq:MS_belief}) computed at time $t$, and $\gamma(t)$ a used-defined reinforcement schedule. In practice we used the same value at each iteration $\gamma(t)\equiv\gamma\sim N_{iter}^{-1}$ where $N_{iter}$ is the number of iterations.

\subsection{Survey Propagation}
In regions of the parameter space in which the 1RSB formalism is more appropriate, one could try to employ an iterative algorithm based on its description. One possibility is Survey Propagation, that in a completely analogous way to the Belief Propagation algorithm, iterates 1RSB equations (described later in \eqref{eq:1rsb_energetic_singleinstance}). %(described later in \eqref{eq:1RSB_xgen} for $x=0$). 
 Survey propagation is complemented with a decimation procedure.

%% file: sections/deg2_results.tex
\section{A simple case: cycle codes}
\label{sec:deg2_results}

We start our analysis with a family of linear systems called cycle codes. 
They correspond to systems of linear equations (in $GF(2)$) in which each variable participates in at most $2$ equations. 
In the graphical representation, a cycle code is a bipartite graph $\cG=(V,F,E)$ in which each variable node $i\in V$ has degree $\le2$.
This particular ensemble has a simple structure that allows to provide exact results. 
In particular, we provide in~\ref{subsec:proof_MS} a rigorous proof of the exactness of Max-Sum solution. 
We also design a greedy optimal (GO) algorithm that is guaranteed to find the optimal solution (see \ref{subsec:optimal_algo}).

\subsection{Comparison of cavity predictions and algorithmic performances on single instances}
\label{subsec:deg2_cavityresults}
\begin{figure}
	\centering
	\includegraphics[width=\columnwidth]{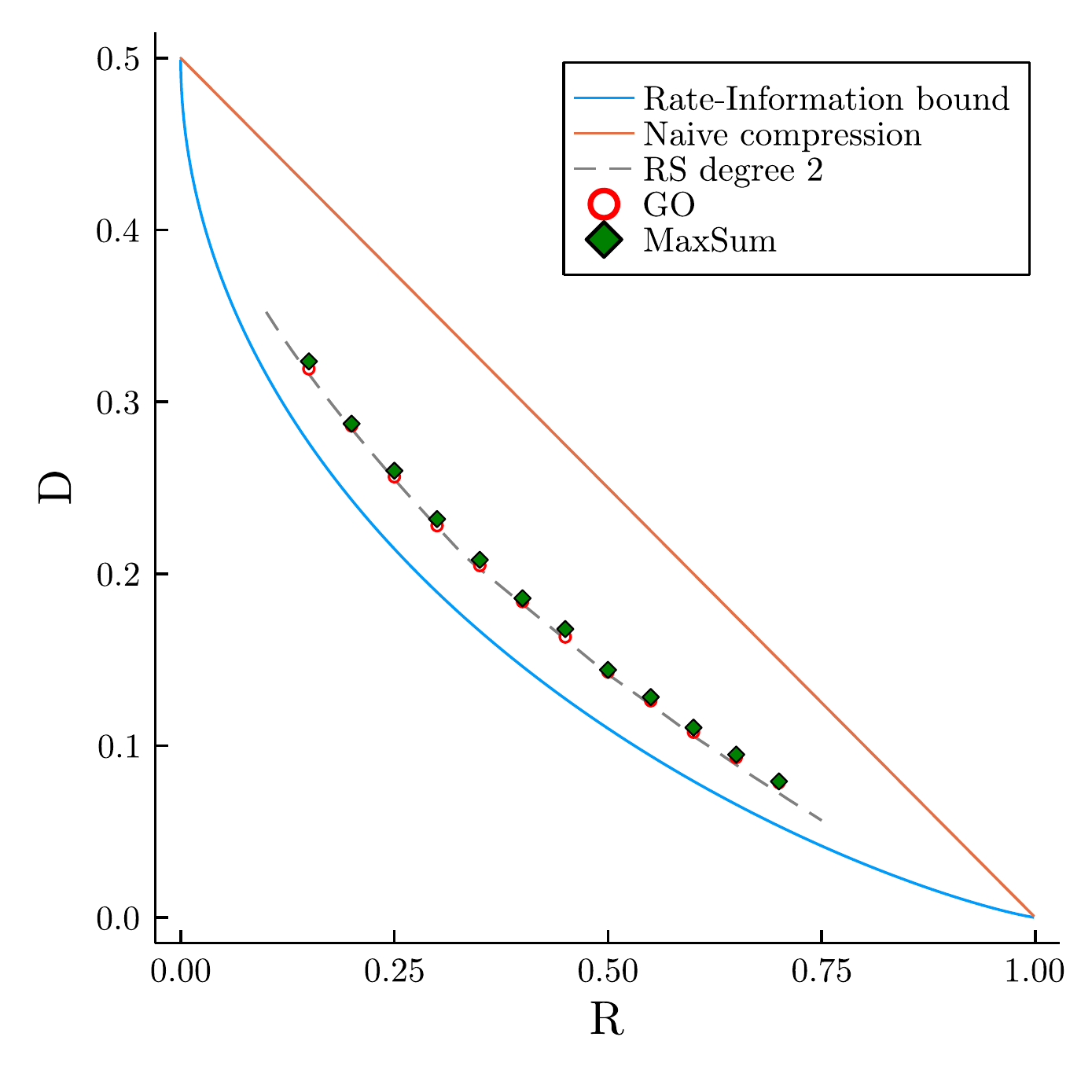}
	\caption{Results for cycle codes: rate-distortion performance for the algorithms GO and Max-Sum with reinforcement on graphs of size $n=1800$, degree profile $\Lambda(x)=x^2$, $P(x)=p_kx^k+p_{k+1}x^{k+1}$. Points are the average over 20 random graphs and source vectors.}
	\label{fig:deg2}
\end{figure}

We focus on a family of random graph ensembles with factor degree profile $P=\{p_k, p_{k+1}\}$, with $k$ a positive integer, and $p_k, p_{k+1}\in[0,1]$ with $p_k+p_{k+1}=1$.
The variable degree profile is $\Lambda=\{\lambda_2=1\}$: each variable node has fixed degree $2$. 
The rate for this family of instances can be expressed as a function of $k$ and $p_{k+1}$:
\begin{align}
\label{eq:rate_deg2}
	R(k,p_{k+1})=1-\frac{2}{k+p_{k+1}}
\end{align}
Varying $k$ and $p_{k+1}$ allows to span the range $R\in[0,1]$.
Fig.~\ref{fig:deg2} shows the performance of the algorithms GO (red circles) and Max-Sum with reinforcement (green squares) for graphs with variables of degree 2, compared to the zero-temperature Replica Symmetric prediction (gray dashed line).

As in the rest of the paper, results in Fig.~\ref{fig:deg2} are presented in the rate-distortion plane, using the framework of lossy compression. 
The blue line corresponds to the exact rate-distortion bound given in equation~(\ref{eq:rate-dist-bound}), i.e. the minimal distortion achievable at a given rate $R$.
The red line corresponds to the distortion achieved with the trivial compression strategy described in the introduction (see~\ref{subsec:compression}).
Note that points relative to Max-Sum have a slightly larger distortion than the results of GO.
This is due to the fact that Max-Sum does not converge on all instances, contrarily to the algorithm GO that provides the exact solution on all instances. 
In case of non-convergence, the strategy adopted was to split the variables in a set of independent and dependent variables, as explained in \ref{subsec:decimation}. After some running time, although Max-Sum algorithm solving (\ref{eq:MS}) has not converged, one fixes the value of the independent variables according to their Max-Sum belief (\ref{eq:MS_belief}), and fixes the dependent variables in order to satisfy the set of linear constraints.

\begin{figure}
	\centering
	\includegraphics[width=\columnwidth]{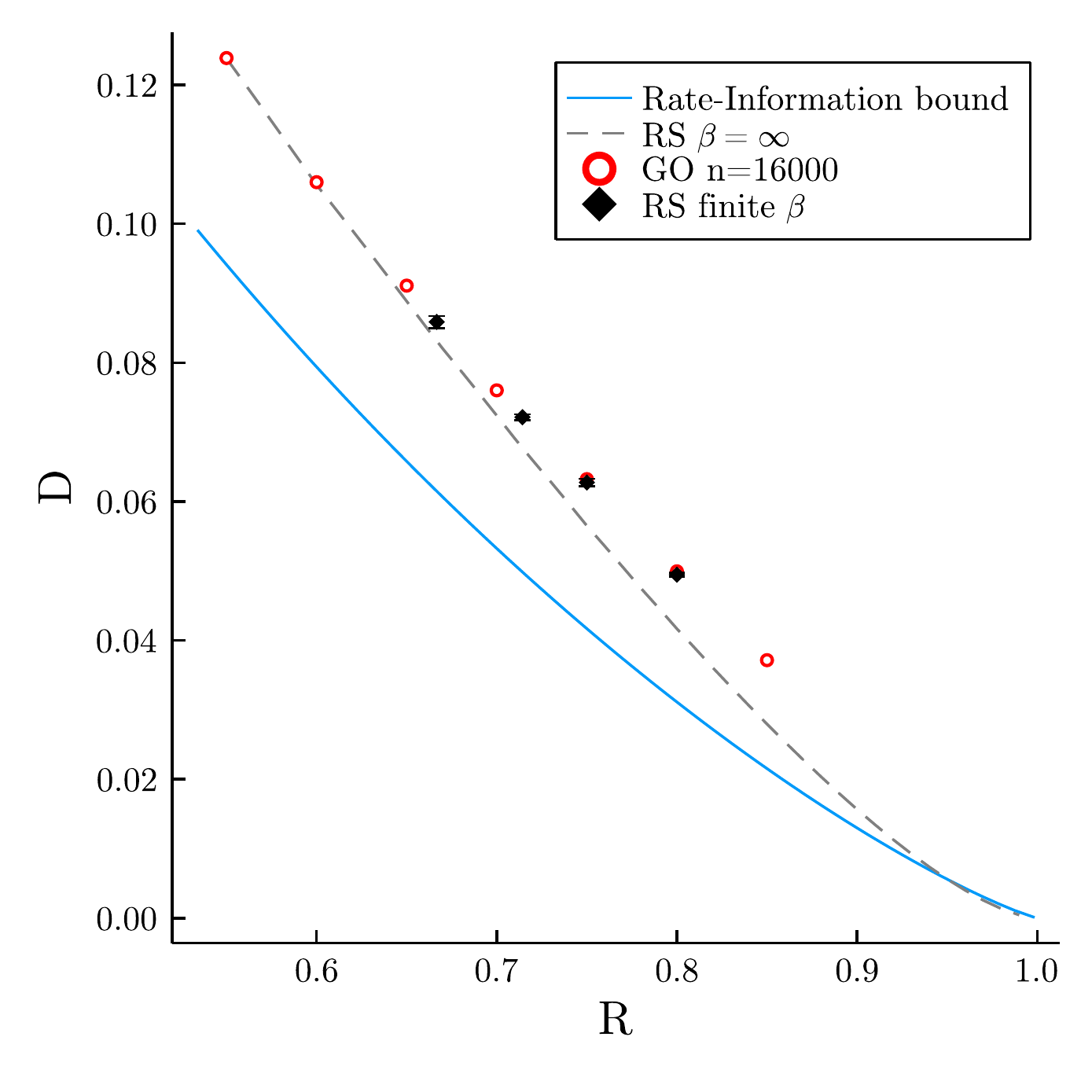}
	\caption{Results for cycle codes: for rates $R>0.6$ there is a discrepancy between the RS zero-temperature prediction (gray dashed line) and the GO results on large instances (red circles). A zero-temperature extrapolation of the RS finite-temperature cavity method (black diamonds) allows to recover the correct distortion.}
	\label{fig:deg2-highrate}
\end{figure}

For rate up to $R=0.6$ (i.e. for $k\leq5$)), we observe a good agreement between the zero-temperature Replica-Symmetric cavity prediction and the results of the algorithms (GO and Max-Sum).
Indeed, we found numerically that the unique solution of the zero-temperature 1RSB equations (\ref{eq:1RSB_energetic}) is the trivial RS solution (\ref{eq:trivial_RS_energetic}) for $R<0.6$, which confirms that we are in the Replica Symmetric phase in this regime. 
For larger rates ($R>0.6$), there are several signs indicating that the zero-temperature RS solution is not correct: first there is a discrepancy between its predicted average distortion and the results obtained with GO on large ($n=16000$) instances (see Fig.~\ref{fig:deg2-highrate}). 
Above $R=0.95$, the distortion computed with the RS ansatz even goes below the exact rate-distortion lower bound.
We were able to obtain a physical solution by considering the finite-temperature cavity method described in appendix~\ref{subsec:entropic_cavity_method} (see Fig.~\ref{fig:deg2-highrate}, black points are a large $\beta$ extrapolation of the RS prediction at finite $\beta$).
We leave for future work a further exploration of this regime, in particular to explain the discrepancy between finite and zero-temperature cavity methods.
It might be that in this regime one cannot exchange the thermodynamic and zero-temperature limit, and therefore that the zero-temperature cavity method is not correct. 
It is also possible that a RSB transition occurs as $k$ increases, however we could not confirm this hypothesis since we encountered some convergence issues when trying to solve numerically the 1RSB equations at zero temperature (\ref{eq:1RSB_energetic}).

\subsection{Exactness of the Max-Sum fixed points}
\label{subsec:proof_MS}
For cycle codes, Max-Sum fixed points correspond to optimal solutions. This can be seen as a consequence of \cite{Wainwright04} (extension of Thm 3 mentioned in the conclusive part), plus a certain property that guarantees that local optimality ensures global optimality for cycle codes. More explicitly, any sub-optimal configuration can be improved by modifying it along a cycle or open path ending on leaves. We will provide however a separate proof for the sake of completeness.

A key hypothesis is that degeneracy on the ground state is removed (by e.g. adding an small random external field on each variable) so that the minimum is unique.

The main tool being used in the proof is the computation tree, whose features are reviewed here, before stating the proof.

\subsubsection{Computation trees}
A computation tree is a loop-free graph obtained from a loopy one.
Here we will follow the notation in \cite{bayati2008max}.
In principle, computation trees can be built for any graph, although here we will focus on factor graphs.

Given a bipartite graph $\mathcal{G}=(V,F,E)$, the idea is to pick a variable node $i_0\in V$ to be the root and then unroll $\mathcal{G}$ around it, respecting the local structure of each node.
Starting from the root, the first level of the computation tree is formed by the root's neighbors.
The second level is made of the first level nodes' neighbors, except the root.
The third level is made of the second level nodes' neighbors, except the ones already connected from above.
Proceeding in this way for a number $k$ of levels produces the level-$k$ computation tree with root $i_0$ called $T_{i_0}^k$.

\input{images/computation_tree.tex}

The concept is best explained through an example, shown in Fig.~\ref{fig:ct}.
Nodes in the original graphs have multiple counterparts in the computation tree.
Sometimes it is useful to refer to those counterparts with the same name as the respective nodes in the original graph.

\subsubsection{Statement and proof} 

The problem of finding the minimum of the distortion or energy $E$ expressed as 
\begin{equation}\label{eq:factorizedenergy}
    \min_{\underline{x}:H\underline{x}=0}E(\underline{x}) = \min_{\underline{x}:H\underline{x}=0} \sum\limits_{i=1}^n{x_i\oplus y_i},
\end{equation}
i.e. the Hamming distance between $\underline{x}$ and the source vector $\underline{y}$.
Consider a MS fixed point on the factor graph $\mathcal{G}$. Call $\{h_{i\to a}, u_{a\to i}\}_{\substack{i\in V\\ a\in F}}$, $\{h_i\}_{i\in V}$ respectively the messages and beliefs and let 
\begin{equation}
g_i = 
\begin{cases}
0 & \text{if $h_i>0$}\\
1 & \text{if $h_i<0$}
\end{cases}
% g_i = \argmax_{x_i}\left(h_i\oplus x_i\right)\quad \forall i\in V
\end{equation}  
be the decision variables.

\paragraph{Claim.}
If there is a unique solution, and if the Max-Sum algorithm has converged, then the set of decision variable corresponds to the optimal solution.

\paragraph{Proof.}
Pick a node $i_0$ and build the computation tree $T_{i_0}^k$ obtained by unrolling the original graph around $i_0$ until there are at least $r$ counterparts of each vertex in $V$ and so that all leaves are variable nodes; $i_0$ will be the root.

Now place messages $\{u_{a\to i}\}$ on the edges of $T_{i_0}^k$ that correspond to the original edges $\{(a,i)\}$ on $\mathcal{G}$. 
Attach to all nodes that are leaves in the computation tree but were not leaves in the original graph, a fictitious factor that sends a message equal to the one flowing out of the leaf. 
By construction, decision variables on the computation tree are now equal to the ones in the original graph.
% Since messages are the same on $\mathcal{G}$ and on $T_{v_0}^k$, so will  beliefs, and therefore decision variables.

Consider now the optimization problem with the same structure of the original one but defined on the computation tree. Messages on $T_{i_0}^k$ constitute a solution for the MS equations: they are naturally fulfilled in the inner part of the tree and imposed on the leaves by the fictitious factors.
Since MS is exact on trees, the assignment $\{g_i\}$ (replicated for each of the counterparts in $T_{i_0}^k$ of each variable in $\mathcal{G}$), is an exact solution for the problem defined on the computation tree.

Now call the optimum for the original problem
\begin{equation}
    \underline{x}^* = \argmin\limits_{\underline{x}:{H\underline x=0}} E(\underline{x}).
\end{equation}
Suppose (absurd) that decision variable $g_{i_0}$ for the root is different from its value in the optimal assignment, $x^*_{i_0}$. Namely, $g_{i_0} = \overline{x^*_{i_0}}$, where $\overline{x}$ is the complement of $x$ under $\gftwo$.
We show that it is always possible to find a path $\mathcal{P}$ on $T_{i_0}^k$ such that complementing every variable along such path results in an improvement in the objective function of the problem defined on the tree. This contradicts the fact that $\{g_i\}$ is an optimum.

The key idea is the following: suppose two vectors $\underline{x}_1,\underline{x}_2$ are both solutions of $H\underline{x}=0$ but differ in the root variable $i_0$.
If $i_0$ is disconnected from the rest of the graph, $\underline{x}_1$ and $\underline{x}_2$ can have all the other bit equal to each other and be two solutions.
If $i_0$ has degree 1, then the single factor attached to it must have at least another incident variable with value different in $\underline{x}_1$ and $\underline{x}_2$ in order for the parity check to be satisfied.
If $i_0$ has degree 2, then the above must be true for both factor  neighbors.
With these observation in mind, let us move to the explicit construction of path $\mathcal{P}$.

To construct path $\mathcal{P}$, start from the root $i_0$, pick any of its factor neighbors, which by construction are at most two, and do the following: look at all the variable nodes incident on the factor and pick one for which the decision variable disagrees with the optimal solution. There will always be at least one in order for the parity-check to be satisfied, as explained before. Include that variable in the path and move on to its other factor if there is any. The process is halted when either a leaf is encountered and the path ends, or the root is found again.

In case the path ends on a leaf, go back to the root and repeat the process in the other direction, or end the path on the root if the root is a leaf.
In case the path ends in a cycle, carry on extending the path repeating the same choice of variables to be included, until the leaves of the computation tree are reached.
Let us stress that the resulting path $\mathcal{P}$ only touches variables which have different values in the solutions on $\mathcal{G}$ and $T_{i_0}^k$.

At this point, the path stemming on both sides from the root can either end on a ``true'' leaf of $T_{i_0}^k$ (one that corresponds to a leaf also on $\mathcal{G}$) or continue until the bottom of the tree, where the fictitious factors are.
We prove our claim in the worst case, where both branches of the path go all the way down, the others follow easily.

Call $\mathcal{P'}$ the projection of $\mathcal{P}$ on $\mathcal{G}$: it may contain cycles.
Again thanks to the fact that no parity check can be left unsatisfied, if $\mathcal{P'}$ is an open path, then its endpoints must be leaves of $\mathcal{G}$.
Call $E_0$ the energy of the optimal configuration $\underline{x}^*$ and $E_0 + \epsilon$ the energy of the first non-optimal one. Further, call $\underline x_{\mathcal{P}}$ and $\underline x_{\mathcal{P'}}$ the indicator functions of paths $\mathcal{P}$ on $T_{i_0}^k$ and $\mathcal{P'}$ on $\mathcal{G}$ respectively. 

For sure, since $\underline{x}^*$ gives a minimum, a transformation $\underline{x}^* \oplus x_{\mathcal{P'}}$ that complements the variables touched by $\mathcal{P'}$ gives a positive shift in energy
\begin{equation}
    E\left(\underline{x}^* \oplus \underline{x}_{\mathcal{P'}}\right) \ge E_0+\epsilon
\end{equation}

Because the energy function \eqref{eq:factorizedenergy} is a sum over functions of single nodes, the shift in energy is only due to the flip of variables in $\mathcal{P'}$.
After the flip, all the touched variables assumed the value they have on $T_{i_0}^k$.
But this means that complementing them on $T_{i_0}^k$ would reverse the shift, thus lowering the energy of the problem defined there by at least $\epsilon$ for each repetition of $\mathcal{P'}$, at least in the bulk.

If $\mathcal{P'}$ is a path, then the same negative shift in energy happens along $\mathcal{P}$ on $T_{i_0}^k$, finding a better optimum than $\{g_i\}$ and thus contradicting the starting point.
If instead, $\mathcal{P}$ goes down all the way to the fictitious factors, the improvement gets multiplied times the number of repetitions of $\mathcal{P'}$, although it might in principle be outbalanced by the change in energy due to the interaction of the one or two leaves $\mathcal{P}$ ends on with their fictitious factors. An upper bound for this changes is given by the maximum absolute difference in messages on the tree
\begin{equation}
    u_{\text{max}} = \max\limits_{(i,a)\in\mathcal{P'}} \vert u_{a\to i}\vert.
\end{equation}
Since there is no limit to the tree's depth, it suffices to repeat the path enough times to be sure that energy will decrease. This amounts to choosing $k$ so that
\begin{equation}
k\epsilon > u_{max}
\end{equation}
Namely,
\begin{equation}\label{eq:k}
    k >\frac{u_{\text{max}}}{\epsilon}
\end{equation}
The same argument can be repeated for all $v_0$'s for which the solution on the computation tree differs from the presumed optimal one.

This completes the proof.

\subsection{An optimal greedy algorithm}
\label{subsec:optimal_algo}
In this section we present the greedy optimal algorithm GO that performs a local search in the energy landscape, decreasing at each time step the energy $E(\underline{x}) =\sum_{i=1}^n x_i\oplus y_i$, by flipping variables in the codeword $\underline{x}\in\mathcal{C}$ in order to obtain a codeword $\underline{x}'$ with lower energy $E(\underline{x}')<E(\underline{x})$. 

For this algorithm all variable nodes should have fixed degree $2$ (if not, one can add an equation to the system involving all leaf variables. This equation is clearly linearly dependent on the other ones, as it is the $GF(2)$ sum of all rows). One considers a slightly simplified factor graph $\mathcal{G'}=(F,E')$ in which the set of new vertices is the set of factor nodes $F$, and the edges $e'=(a_1,a_2)\in E'$ are linking two vertices $a_1,a_2\in F$ through the variable node $i\in V$ such that $\partial i=\{a_1,a_2\}$. For any codeword $\underline{x}={}^t(x_1,\dots,x_n)$ and source vector $\underline{y}={}^t(y_1,\dots,y_n)$, one defines a weight function $w_{(\underline{x},\underline{y})} : E'\to\{-1,1\}$ on the graph $\mathcal{G'}$ as follows. The edge $e_i'\in E'$ passing through the variable node $i$ associated to the $i^{\rm th}$ boolean component has weight $w_{(\underline{x},\underline{y})}(e_i')=+1$ if the components of the source vector $\underline{y}$ and of the current codeword $\underline{x}$ coincides: $x_i=y_i$, and $w_{(\underline{x},\underline{y})}(e_i')=-1$ otherwise. One then look for a negative cost cycle $\mathcal{L}$, using the algorithm presented in \cite{GuMaSuLa09}. Let $\underline{x}_{\mathcal{L}}$ be the indicator function on the negative cost cycle $\mathcal{L}$. Flipping the variables belonging to $\mathcal{L}$ results in a new codeword $\underline{x}'=\underline{x}\oplus \underline{x}_{\mathcal{L}}$ with a strictly smaller energy.

The algorithm starts with a given codeword $\underline{x}_0\in\mathcal{C}$ which may not be the optimal codeword. Then at each time step $t$ it finds a negative cost cycle $\mathcal{L}_t$ for the weight function $w_{(\underline{y},\underline{x}_t)}$, and flip the variables in the cycle to get a lower energy state: $\underline{x}_{t+1}=\underline{x}_t\oplus \underline{x}_{\mathcal{L}_t}$. One repeats this procedure until convergence, i.e. when the difference between energies $\Delta E=E(\underline{x}_{t+1})-E(\underline{x}_t)$ becomes zero (up to numerical precision). Finding a negative cost cycle can be done efficiently \cite{Gu09}.

%% file: images/computation_tree.tex
\begin{figure}[h]
    \includegraphics[width=0.49\columnwidth]{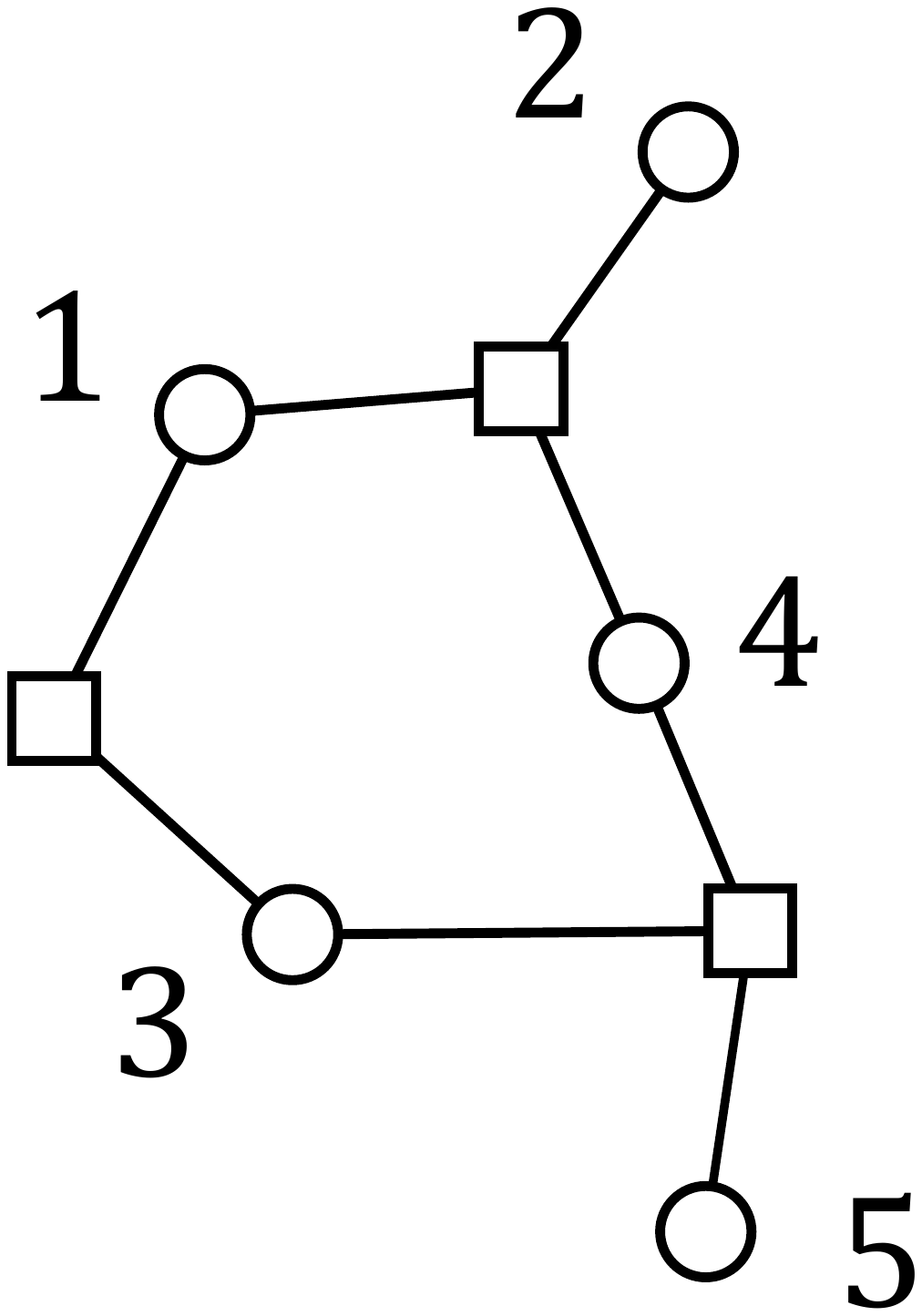}
    \includegraphics[width=0.49\columnwidth]{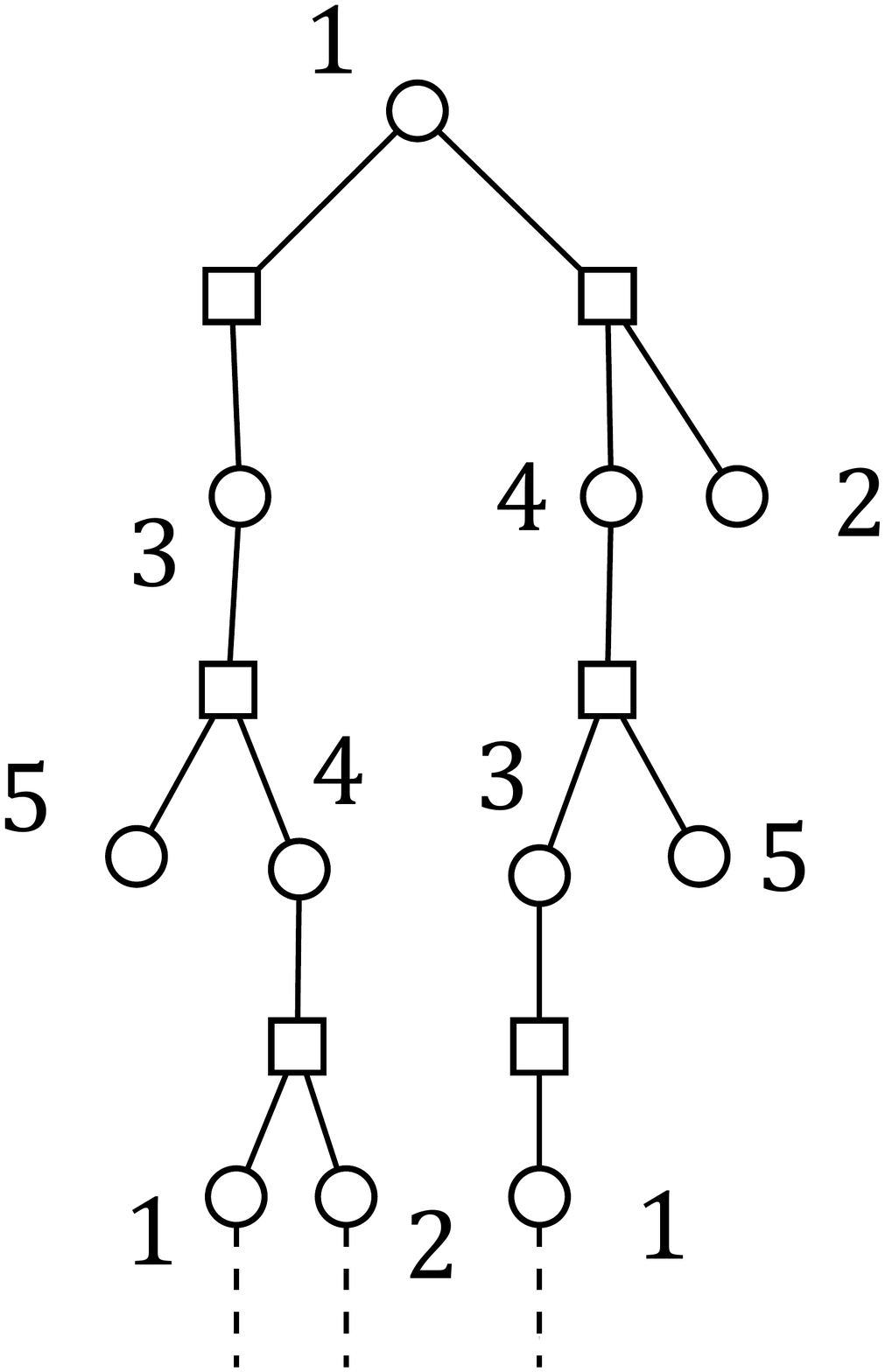}
    \caption{A graph (left) and its computation tree rooted at variable $1$ (right). Variables are indicated by circles and factors by squares. The dashed lines at the bottom in the right graph indicate that the tree can grow to arbitrary depths}
\label{fig:ct}
\end{figure}

%% file: sections/deg3_results.tex
\section{Moving to higher degrees}
\label{sec:deg3_results}
We have shown in the previous section that the constrained optimization problem CVP on cycle codes (i.e. in which variable nodes have degree at most $2$) could be solved exactly.
In this section we study random graph ensembles in which variable nodes have higher degree. 
We show that moving to these ensembles allows to reach a smaller minimal energy. 
We focus on random graph ensembles with variable degree profile $\Lambda=\{\lambda_2,\lambda_3\}$, i.e. with a fraction $\lambda_2$ of variable nodes with degree $2$, and a fraction $\lambda_3=1-\lambda_2$ of variable nodes with degree $3$.
The factor degree profile is $P=\{p_3=1\}$, i.e factor nodes have fixed degree $3$. 
This is the simplest choice providing a non-trivial phase diagram.
In this ensemble the compression rate can be expressed in terms of the fraction of degree $3$ variables:
\begin{align}
\label{eq:rate_deg23}
	R(\lambda_3) = \frac{1-\lambda_3}{3}\in[0,1/3]
\end{align}
 We will compare this random graph ensemble to cycle codes, where the variable degree profile is $P=\{p_3=1\}$, $\Lambda=\{\lambda_1, \lambda_2\}$ (i.e. there is a fraction $\lambda_1$ of degree-$1$ variables, and a fraction $\lambda_2=1-\lambda_1$ of degree-$2$ variables). In this ensemble we express the rate in terms of the fraction of degree $1$ variables:
\begin{align}
\label{eq:rate_deg12}
    R(\lambda_1) = \frac{1+\lambda_1}{3}\in[1/3,2/3]
\end{align}

\begin{figure*}
	\centering
	\includegraphics[width=\textwidth]{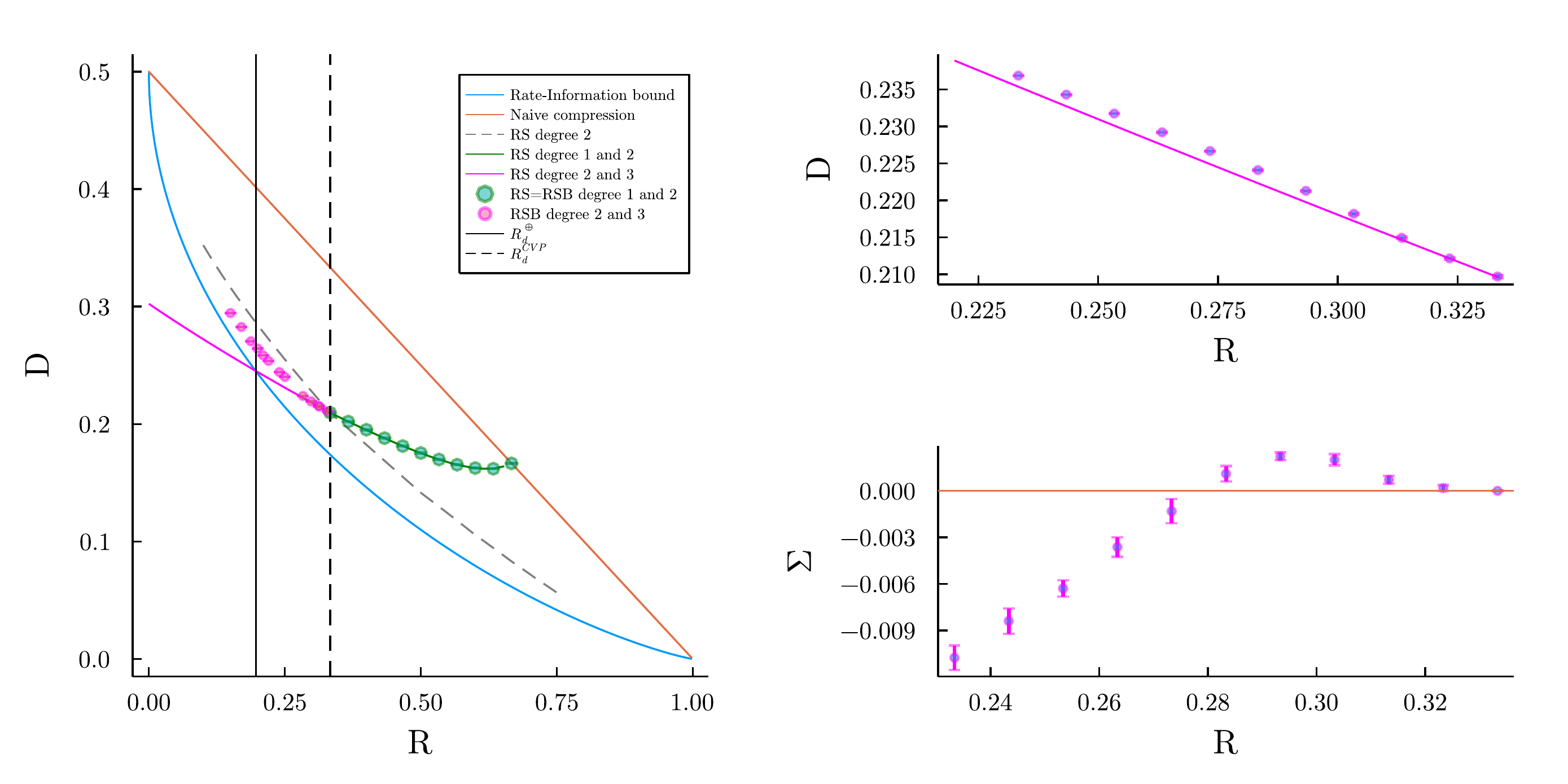}
	\caption{Left panel: zero-temperature cavity prediction of the minimal distortion. In green: for variable nodes of degree $1$ and $2$ (\ref{eq:rate_deg12}), in pink: for variable nodes of degree $2$ and $3$ (\ref{eq:rate_deg23}). 1RSB points are obtained by optimizing on the Parisi parameter $y$, see ~\ref{subseq:instability_energetic_solution}. Plain lines correspond to the RS prediction, circles correspond to the 1RSB prediction. The gray dashed line corresponds to the RS prediction for the ensemble studied in the previous section (\ref{eq:rate_deg2}). The two vertical lines corresponds to the clustering transitions occuring for the CVP (at $R_d^{\rm CVP}=1/3$, dashed line) and for the XORSAT problem (at $R_d^\oplus=0.197$, plain line). Right panel: detail close to $R=1/3$ obtained at fixed $y=1.0$ for the minimal distortion (top) and the complexity (bottom) showing that the clustering transition is continuous.}
	\label{fig:cavity-predictions-deg123-yopt}
\end{figure*}

\subsection{Results from the cavity method}

Fig.~\ref{fig:cavity-predictions-deg123-yopt}, left panel shows the results of the zero-temperature cavity method, under the RS formalism (plain lines) and the 1RSB formalism (circles). 
For the ensemble with variable nodes of degree $1$ and $2$ corresponding to rate (\ref{eq:rate_deg12}) (in green), we see that the RS and 1RSB predictions are the same. 
Indeed for this ensemble the unique 1RSB solution that we found was the trivial RS solution.
For this graph ensemble, the minimal distortion achievable is larger than the one for the graph ensemble studied in the previous section with rate (\ref{eq:rate_deg2}), which is also reported in Fig.~\ref{fig:cavity-predictions-deg123-yopt} (gray dashed line).
It is more interesting to look at the graph ensemble in which variable nodes have degree $2$ and $3$ corresponding to (\ref{eq:rate_deg23}) (in pink).
The RS prediction (pink line) is clearly unphysical, because at small rates it goes below the rate-information bound (blue line). 
The 1RSB formalism is needed to give a reliable prediction of the minimal distortion for this ensemble: pink circles correspond to the 1RSB solution obtained at Parisi parameter $y=y^{\rm opt}(R)$, see the discussion in~\ref{subseq:instability_energetic_solution}.
We see that it enters in a 1RSB phase as soon as there is a positive fraction of degree $3$ variables $\lambda_3>0$ (see in particular the details close to $\lambda_3=0$ in the right panel of Fig.~\ref{fig:cavity-predictions-deg123-yopt}). 
The clustering transition occurs therefore at $R_d^{\rm CVP}=1/3$, and is represented by the vertical dashed line in Fig.~\ref{fig:cavity-predictions-deg123-yopt}, left panel. 
This 1RSB prediction is confirmed by a finite size analysis presented in subsection \ref{subseq:exact_enum}. 
We could reach the physical solution down to rate $R=0.15$ (left-most pink circle). 
For smaller rates, we could not find a physical solution.
We give more details on the numerical resolution of the 1RSB equations in the next section (see~\ref{subseq:instability_energetic_solution}), in particular on the difficulties encountered for small rates.

The vertical plain line indicates the rate at which the dynamical transition occurs for the XORSAT problem: $R_d^\oplus =0.197$, which corresponds to the clustering transition associated with the measure (\ref{eq:prob}) at $\beta=0$. 
In~\ref{subsec:finite_T_study}, we compute the clustering transition $R_d(\beta)$ for finite values of $\beta$, thus making the interpolation between the clustering threshold $R_d^\oplus$ at $\beta=0$ and $R_d^{\rm CVP}$ at $\beta=\infty$ (see Fig.~\ref{fig:entropic-dyn-transition}).
There is therefore a range of the rate $R\in[R_d^\oplus,R_d^{\rm CVP}]$ for which the constrained optimization problem is in a 1RSB phase, while the underlying XORSAT problem defining the set of constraints is Replica-Symmetric.
This situation is interesting because it means that the structure of the set of constraints is not enough to describe the complexity of the constrained optimization problem.
In the range $[R_d^\oplus,R_d^{\rm CVP}]$, the set of allowed configurations (defined as the solution set of the XORSAT instance) is rather well-connected, yet, and despite the simplicity of the optimization function (\ref{eq:energy}), the optimization problem is in a glassy phase: the energy landscape presents many local minima that are separated by free-energy barriers. 
A similar situation has been recently encountered in other high-dimensional constrained optimization problems, see \cite{UrbaniSlocchi21} for a study of the optimization of a quadratic function where the constraints are modeled by a perceptron constraint satisfaction problem.
Moreover, we observe that the RS/1RSB transition is continuous, see Fig.~\ref{fig:cavity-predictions-deg123-yopt}, right panel.
In contrast, the RS/1RSB transition for the underlying constrained satisfaction problem (the XORSAT problem) is a random first-order transition.

The physical scenario we have found in this constraint satisfaction problem can be summarized as follows: (i) the set of allowed solutions is well connected until it undergoes a random first order transition (RFOT) and become clustered for $R<R_d^\oplus$, (ii) however, already for $R<R_d^{\rm CVP}$, the application of an external field in a random direction induces a continuous transition.
This is reminiscent of what happens in the $p$-spin model \cite{crisanti1992sphericalp}, where the discontinuous phase transition becomes continuous under the application of an external field.
Let us give a simple intuition of the structure of solutions in this problem.
The most abundant solutions are well connected for any $R>R_d^\oplus$. However, this strong connectedness is no longer true as soon as we put a bias in any direction and we concentrate the measure on a subset of solutions: leaving the region where the most abundant solutions live, for any $R<R_d^{\rm CVP}$, the space of solutions acquires a non trivial structure that in turn induces a continuous phase transition and requires the replica symmetry to be broken in order to describe correctly this non trivial structure.
Eventually at $R_d^\oplus$ also the most abundant solutions acquire a non trivial structure (actually undergo a clustering transition) and this is likely to have important consequences for algorithms.
Indeed, while a problem undergoing a continuous transition can be well approximated by polynomial algorithms, we expect a RFOT to induce a much more serious algorithmic barrier. We explore algorithmic consequences in the next subsection.

\subsection{Algorithmic results}

\begin{figure}
	\centering
	\includegraphics[width=\columnwidth]{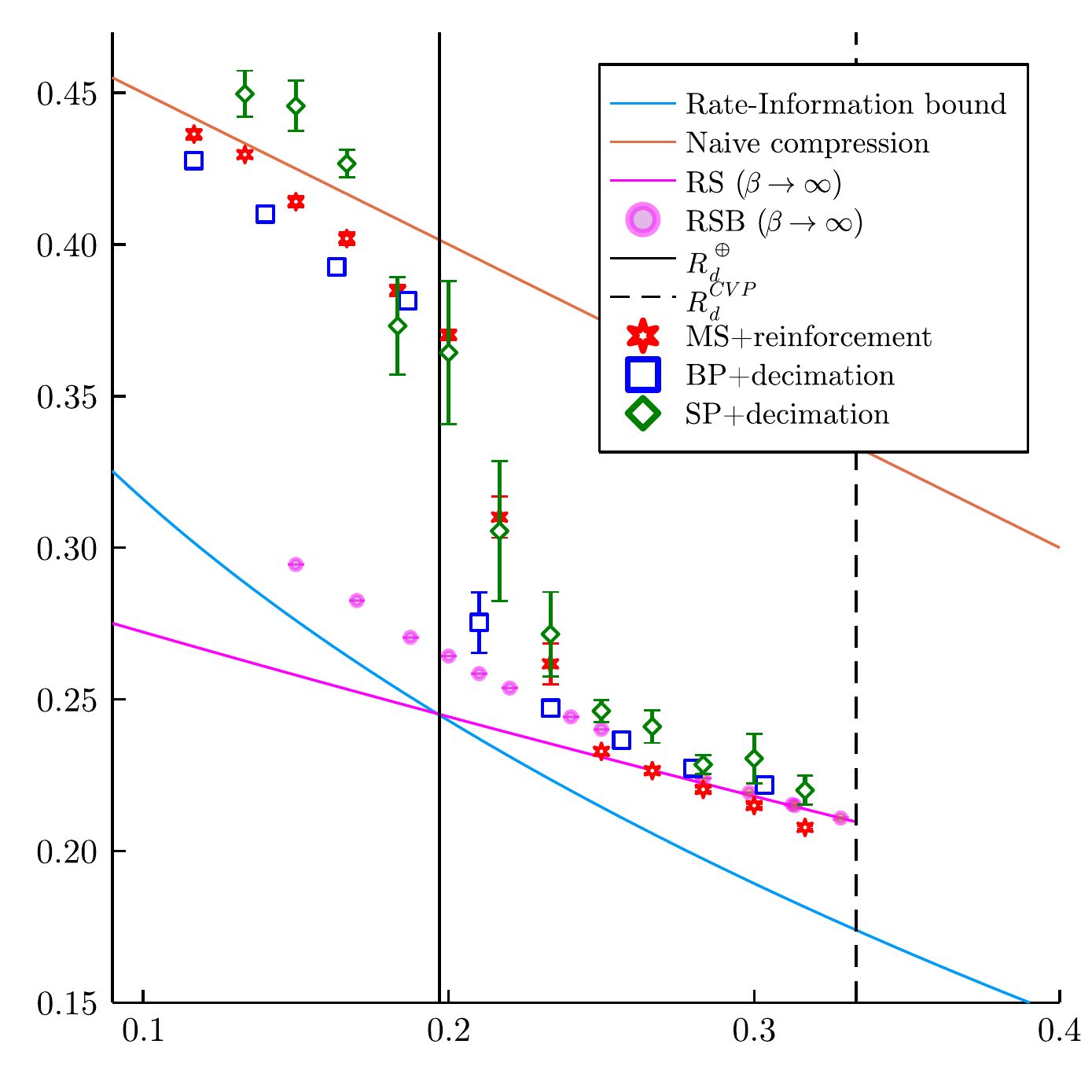}
	\caption{Rate-distortion performance for Max-Sum with reinforcement (red stars), for Belief Propagation with decimation at $\beta=3$ (blue squares), and Survey Propagation at $y$ maximizing the 1RSB free energy (green diamonds), on graphs of size $n=1800$, degree profile $\Lambda=\{\lambda_2, \lambda_3\}$, $P=\{p_3=1\}$. Points are the average over 15 random graphs and source vectors.}
	\label{fig:bp_dec}
\end{figure}

In this subsection, we report the results obtained with the algorithms described in section \ref{sec:definitions}.
Fig.~\ref{fig:bp_dec} shows the results of Max-Sum with reinforcement (red stars), Belief-Propagation with decimation at finite inverse temperature $\beta=3$ (blue squares), and Survey-Propagation with decimation (green diamonds) at $y=y^{\rm opt}(R)$ maximizing the 1RSB free-energy, see the discussion in~\ref{subseq:instability_energetic_solution}. 
The result of the zero-temperature cavity method within the 1RSB ansatz is also reported (pink circles). 
For rates in the range $[0.25, 1/3]$ there is good agreement between the cavity prediction and the algorithmic results. 
As the rate decreases, one observes a jump toward solutions with higher distortion found by the three algorithms, while the cavity method predicts a smaller distortion.
This decrease in performance arises around the XORSAT dynamical transition occurring at $R_d^\oplus=0.197$.
In the clustered regime $R<R_d^\oplus$, none of the three algorithms is able to find the optimal solution. 

This result is interesting, because the algorithmic transition does not match with the phase transition associated to the constrained optimization problem that we found at $R_d^{\rm CVP}=1/3$, instead it matches with the clustering transition for the XORSAT problem that models the constraints (and which is not related to the optimization function).
A possible explanation for the fact that algorithms perform well in the range $[0.25,1/3]$ and undergo a dramatic algorithmic transition only approaching $R_d^\oplus=0.197$ is the following.
As long as $R>R_d^\oplus$ the most abundant solutions are well connected (they undergo a clustering transition only at $R_d^\oplus=0.197$) and the phase transition induced by the external field (the linear function to be optimized) is continuous \footnote{We stress than the problem in this region probably undergoes a full replica symmetry breaking (FRSB) transition, but we are able to perform only a 1RSB computation, that should approximate closely the actual solution.}
In this situation we expect optimizing algorithms to perform in an efficient way: on the one hand the space of solutions is well connected by passing through the most abundant (although not very optimized) solutions and on the other hand a continuous phase transition induces correlations that can be well approximated by polynomial algorithms. For these reasons, we expect smart optimizing algorithms, like the ones we have used, to perform reasonably well above $R_d^\oplus$.
Obviously the performances of these algorithms start degrading already above $R_d^\oplus$, because approaching the clustering transition at $R_d^\oplus$ the structure of solutions starts acquiring a sponge-like topology, that will eventually break up in separated cluster at $R_d^\oplus$. When the topology is sponge-like, with tiny corridors connecting the regions that will become soon clusters, the application of the external field can have dramatic effect, effectively anticipating the clustering transition.

Eventually for $R<R_d^\oplus$ the most abundant solutions are clustered and moving between solutions becomes very difficult, effectively inducing large algorithmic barriers. In this regime the effects of the RFOT are manifest and all the algorithms get stuck in solutions of very large distortion. This reminds the threshold energy phenomenon, well known in glassy models and hard optimization problems \cite{ricci2010being}.

A last important comment about the connection between the dynamical behavior of these smart algorithms and the thermodynamic phase diagram of the problem is about the possibility that smart algorithms do not converge on solutions that dominates the equilibrium measure. This has been already observed in constraint satisfaction problems \cite{dall2008entropy} and in the binary perceptron problem \cite{braunstein2006learning}.
The finite-temperature study reported in section~\ref{subsec:finite_T_study} predicts that the clusters in this problem are point-like, i.e.\ they are made of only one solution, but these solutions should be very hard to find by algorithms.
It is therefore more likely that the solutions found by message-passing algorithms in Fig.~\ref{fig:bp_dec} belong to atypical, large clusters.
We conjecture that these clusters are sub-dominant and thus not described by the cavity method, but yet are the relevant clusters from an algorithmic point-of-view, since they are made of solutions that are more accessible for algorithms. 

\subsection{Exact enumeration}
\label{subseq:exact_enum}

\begin{figure}
    \centering
    \includegraphics[width=\columnwidth]{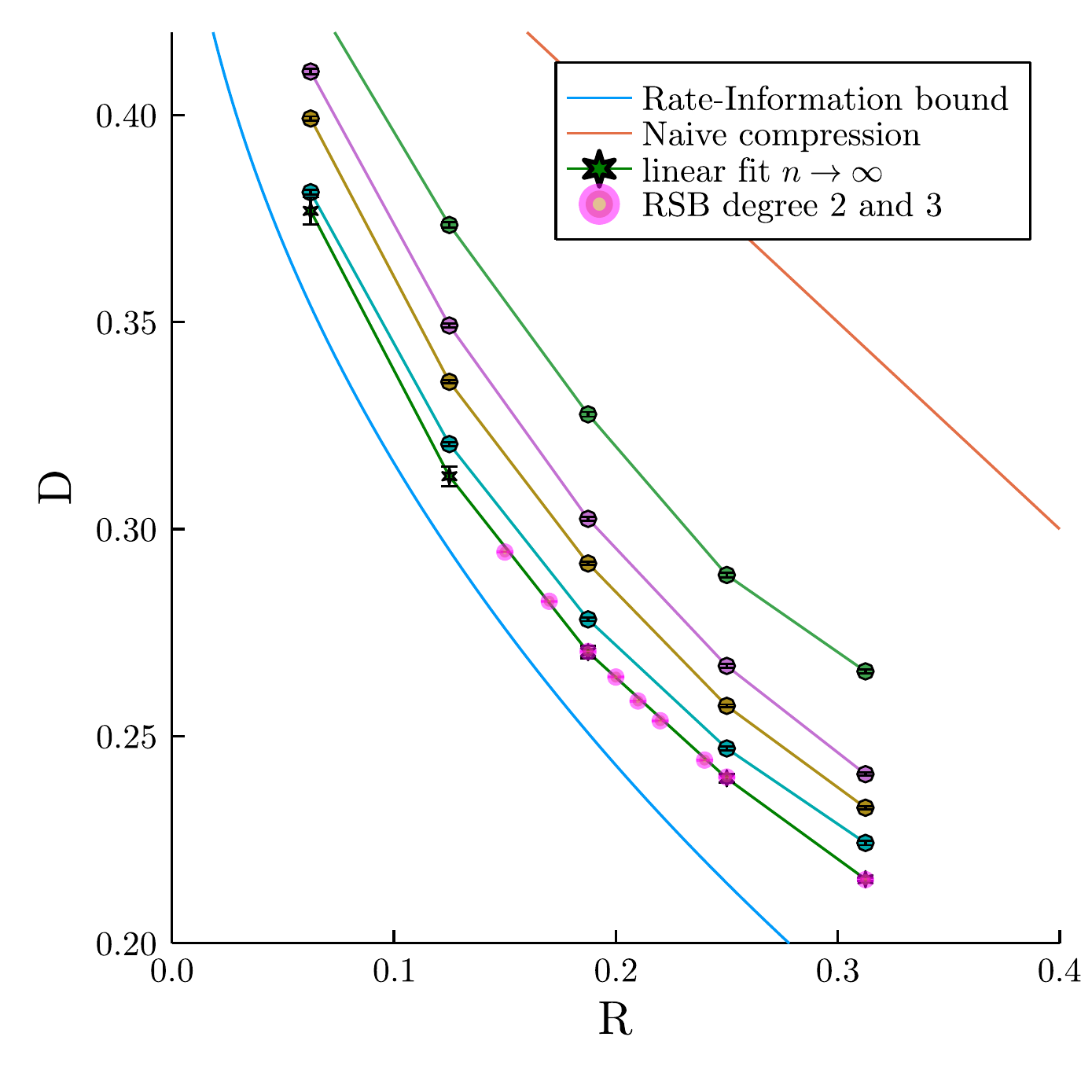}
    \caption{Exact enumeration. The four top lines with points correspond to (from top to bottom) $n=16, 32, 48, 96$. Further down, the $n\to\infty$ predictions are made by means of a linear fit in $1/n$. In pink: prediction of the zero-temperature 1RSB cavity method. }
    \label{fig:exact_enum}
\end{figure}

Given that no linear-time message-passing algorithm can approach the optimal distortion predicted by the RSB cavity method at zero temperature, we need a different numerical approach to convince the reader that the analytical prediction based on the cavity method is actually meaningful.
We have performed an exact enumeration for small sizes and a finite size study of the random graph ensemble with degree profile $\Lambda=\{\lambda_2,\lambda_3\}$, $P=\{p_3=1\}$ to compare with the results of the zero-temperature 1RSB cavity method.
Fig.~\ref{fig:exact_enum} shows the exact results for sizes $n\in\{16, 32, 48, 96\}$. 
The results are averaged over several instances drawn at random from the random graph ensemble $\mathbb{G}_n(\Lambda, P)$. 
For each instance, the solution set of the associated XORSAT instance is computed exactly, and the solution with minimal distortion is extracted.
A linear extrapolation is done in the large size limit, which is in good agreement with the 1RSB cavity prediction.
This exact enumeration procedure allows to give predictions for rates smaller than $R=0.15$, below which the 1RSB cavity method does not provide a physical solution.
We remind that the physically correct solution probably requires the breaking of the replica symmetry infinite times (FRSB), so instabilities in the 1RSB solution at very small rates do not come as a surprise.

%% file: sections/cavity_method_results.tex
\section{A more detailed picture of the phase diagram}
\label{sec:more_detailed_v}

In this section, we give more details on the results obtained with the cavity method, in particular for the random graph ensemble with degree profiles $\Lambda=\{\lambda_2,\lambda_3\}$, $P=\{p_3=1\}$.

\subsection{Instability in the zero-temperature solution}
\label{subseq:instability_energetic_solution}

\begin{figure*}
	\centering
	\includegraphics[width=\textwidth]{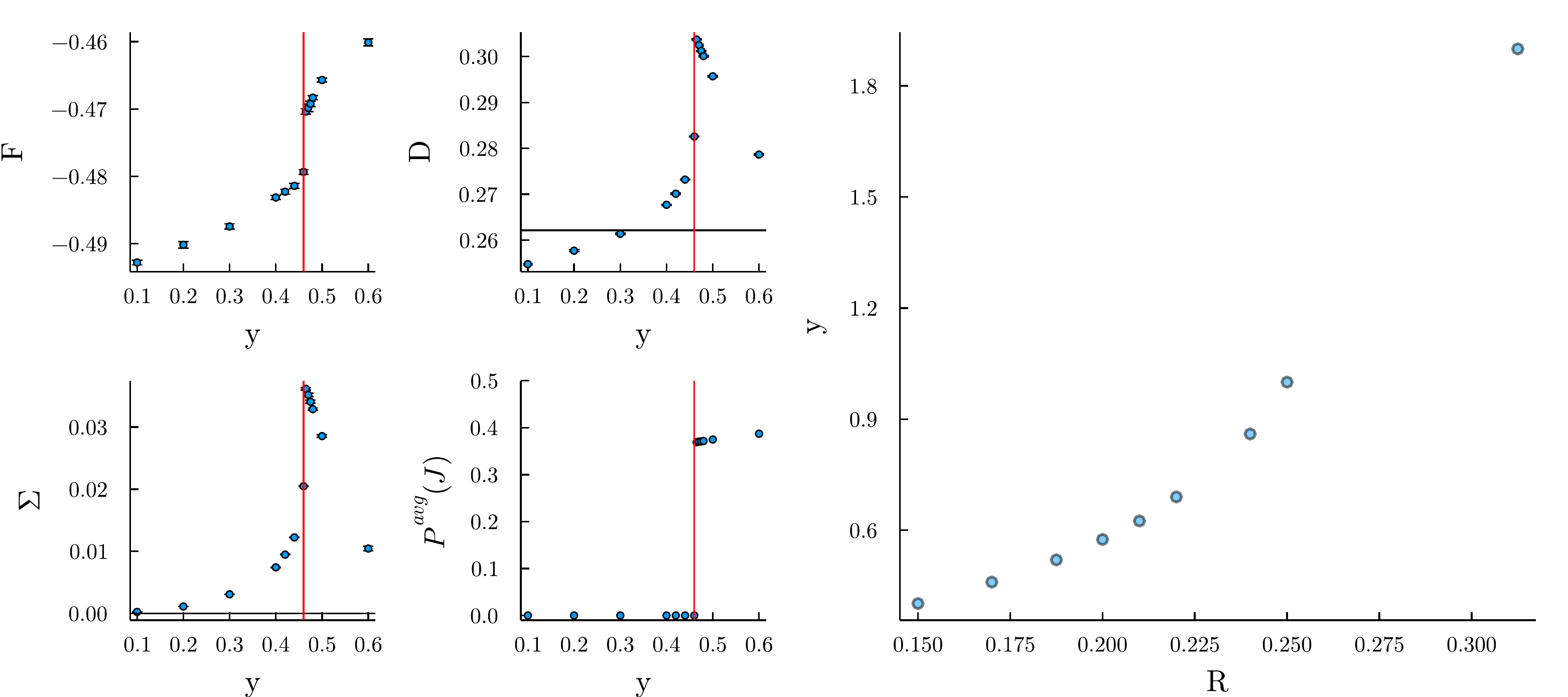}
	\caption{Left panel: $y$ dependence of the RSB solution for $R=0.17$: top left: 1RSB free-energy $F_e^{\rm 1rsb}$, top right: internal distortion $D_{\rm int}(y)$, the horizontal black line indicates the rate-distortion bound $D_I(R)$, bottom left: complexity $\Sigma_e(y)$, bottom right: weight $P^{\rm avg}(J)$. Right panel: optimal value of $y$ computed for several rates. Population size is $5\cdot 10^5$ and $J=20$.}
	\label{fig:y_study}
\end{figure*}

The results of the zero-temperature cavity method (see Fig.~\ref{fig:cavity-predictions-deg123-yopt}) have been obtained by taking simultaneously the limit $\beta\to \infty$ for the inverse temperature and $x\to 0$ for the Parisi parameter, with a finite value for $y=\beta x$.
As explained in the appendix (\ref{subsec:en_cavity_method}) we used the softened measure (\ref{eq:prob_J}) for the numerical resolution of the zero-temperature cavity equations, as it allowed us to represent populations of Max-Sum messages as finite vectors of size $2J+1$.
In the large $\beta$ limit, the softened measure (\ref{eq:prob_J}) concentrates on configurations minimizing the energy (\ref{eq:MinNrjcutoffJ}).
In the 1RSB phase, the set of these configurations is split into an exponential number of clusters separated by free-energy barriers.
Since $x\to 0$, all the clusters are weighted identically (independently of their size).
The choice of the value of $y$ to describe correctly the cluster decomposition is delicate.
Following the seminal work \cite{MezardParisi03}, one should compute the 1RSB free-energy $F_e^{\rm 1rsb}$ defined in (\ref{eq:1rsb_energetic_potential}) and maximize it over $y$. 
In practice, for each value of the rate plotted in Fig.~\ref{fig:cavity-predictions-deg123-yopt}, we studied the $y$-dependence of the solution of the zero-temperature 1RSB equations (\ref{eq:1RSB_energetic}).
Fig.~\ref{fig:y_study}. shows the study of the $y$-dependence for rate $R=0.17$ (left panel), the vertical red line indicates the optimal value $y^{\rm opt}(R)$. 
The right panel shows the optimal value of $y$ as a function of the rate.
For each value of the rate, from the solution of the 1RSB equations we computed the 1RSB free-energy $F_e^{\rm 1rsb}(y)$ and the zero-temperature complexity $\Sigma_e(y)$ (see equation (\ref{eq:complexity_energetic}).
We also computed the internal distortion $D_{\rm int}(y)$, from the internal energy as $D_{\rm int}(y) = 2U_{\rm int}(y)-1$ (see equation (\ref{eq:internal_energy})).
For the appropriate choice of $y$, $D_{\rm int}(y)$ gives a prediction for the minimal distortion.
Finally, we also computed the averaged distribution of cavity fields
\begin{align}
	P^{\rm avg}(h) = \int\dd\cP^{\rm 1rsb}(P)P(h)
\end{align}
From this study one deduces the optimal value of $y$ for each rate $y_{\rm opt}(R)$, as the value maximizing the free energy $F_e^{\rm 1rsb}(y)$, under the following constraints: the complexity is positive $\Sigma_e(y)>0$, the internal distortion is above the Rate-Distortion bound $D_{\rm int}(y)>D_I$, with $D_I$ solution of equation~(\ref{eq:rate-dist-bound}), and the averaged distribution $P^{\rm avg}$ has a zero weight in $\pm J$. 
To justify this last constraint, let us recall the interpretation of the Max-Sum beliefs $h_i$ as the difference between the ground-state energy obtained when the variable $\sigma_i$ is flipped: $h_i=E_i^{\min}(+)-E_i^{\min}(-)$.
If $h_i=\pm J$, when the variable $i$ is flipped, in order to obtain a solution with minimal energy one can either rearrange $O(J)$ other variables, or one can violate a constraint (see the definition of the soft energy function~\ref{eq:MinNrjcutoffJ}), thus resulting in a configuration that is not satisfying all the constraints. We therefore discard solutions having a non-zero weight in $\pm J$, as they could correspond to configurations that are not solution of the XORSAT problem.

From the study of rate $R=0.17$, we see that $F_e^{\rm 1rsb}(y)$ increases smoothly as $y$ increases, until $y=0.46$ where there is a jump toward a solution with $P^{\rm avg}(\pm J)>0$.
Since at $y=0.46$ all the constraints mentioned above are satisfied, we therefore obtain $y^{\rm opt}(R=0.17)=0.46$
As the rate decreases, this $y$-study becomes more and more difficult because this transition becomes more and more pronounced, in particular the jump in $F_e^{\rm 1rsb}(y)$ increases thus lowering the accuracy in the determination of $y^{\rm opt}$.
For rate smaller than $R=0.15$ we could not find a value of $y$ satisfying all the constraints (positive complexity, distortion above $D_I$, and zero-weight in $J$).
Since we did not find a physical 1RSB solution despite our efforts, we conjecture the existence of a transition toward a 2RSB or even full RSB phase at smaller rates. We leave this investigation for future work.

\subsection{A finite temperature study}
\label{subsec:finite_T_study}

\begin{figure*}
	\centering
	\includegraphics[width=\textwidth]{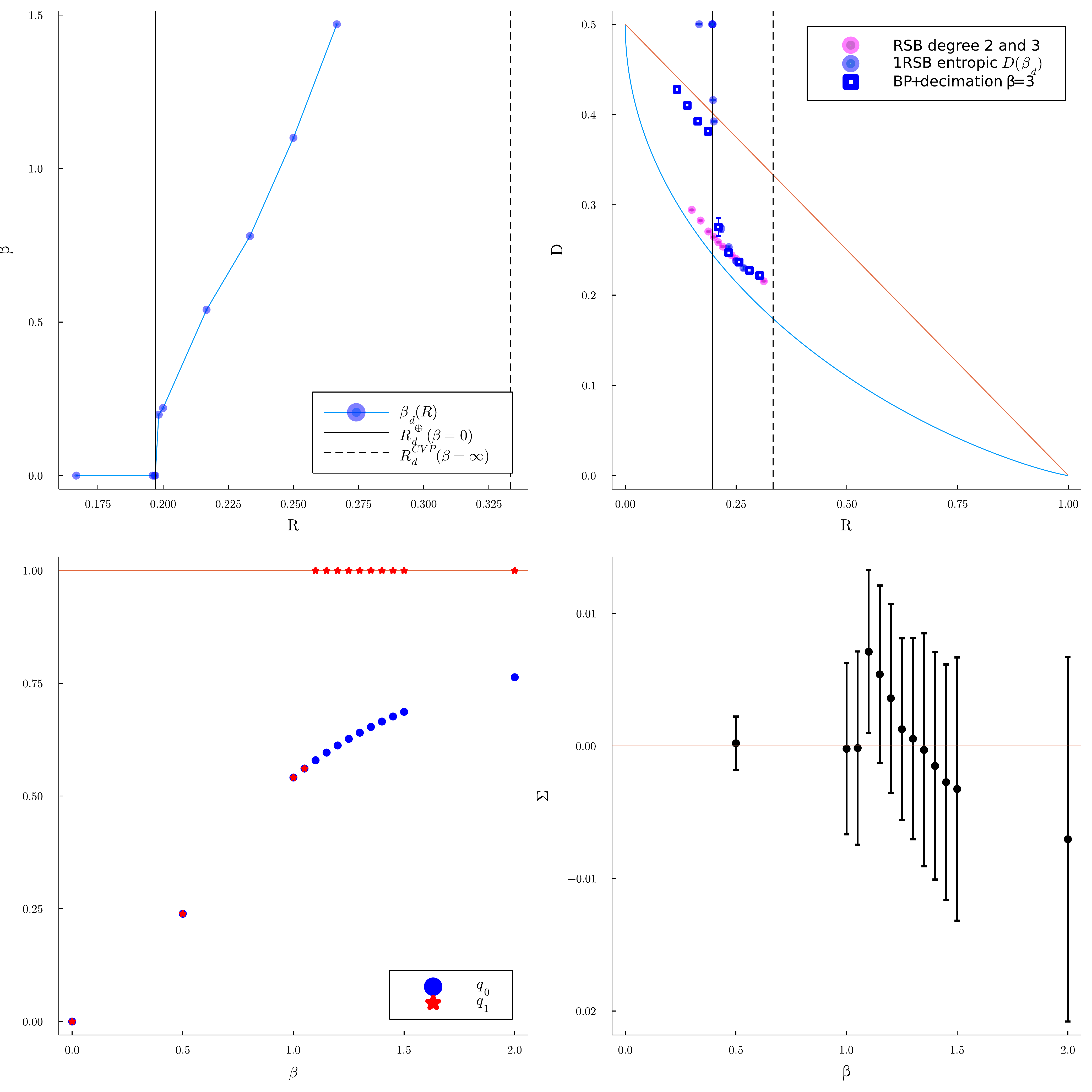}
	\caption{Top left panel: Dynamical (RS/RSB) threshold value $\beta_d(R)$ observed with the finite-temperature method. Top right panel: finite-temperature (blue points) and zero-temperature (pink points) predictions for the minimal distortion, against BP performances at $\beta=3$ (blue squares). The finite-temperature prediction is made at the clustering transition $\beta=\beta_d$. Bottom left panel: Inter-state overlap $q_0$ (blue points) and intra-state overlap $q_1$ (red stars) as $\beta$ increases, at fixed rate $R= 0.25$, showing the clustering transition at $\beta_d=1.1$. Bottom right panel: complexity as $\beta$ increases, at fixed rate $R= 0.25$, showing the clustering transition at $\beta_d=1.1$ and the condensation transition at $\beta_c\approx1.3$.}
	\label{fig:entropic-dyn-transition}
\end{figure*}

In this section, we perform a finite $\beta$ study of the properties of the measure $\mu(\us)$ defined in (\ref{eq:prob}), using the finite-temperature cavity method described in appendix~\ref{subsec:entropic_cavity_method}. 
The objective is to explain the results obtained in the previous section. We saw indeed with the zero-temperature cavity method that the optimization problem (obtained in the large $\beta$ limit) is in a 1RSB phase as soon as $R<R_d^{\rm CVP}=1/3$ (i.e.\ as soon as $\lambda_3>0$).
However, we know that the underlying XORSAT problem (described by the measure (\ref{eq:prob}) at $\beta=0$) is in a RS phase down to $R_d^\oplus=0.197$.
We therefore computed the clustering transition $\beta_d(R)$ line in the $R$-$\beta$ plane, see Fig.~\ref{fig:entropic-dyn-transition} (top left panel), to make the interpolation between $\beta=0$ and $\beta\to\infty$.
What we are finding via this computation is that for any $R<R_d^{\rm CVP}=1/3$ increasing $\beta$ soon or later the measure undergoes a phase transition.

The top right panel shows the finite-temperature RS prediction of the minimal distortion $D_{\min}(\beta)$ computed at the clustering transition $\beta=\beta_d(R)$. 
For $R$ large enough (i.e.\ for $R\gtrsim 0.25$), the finite-temperature prediction $D_{\min}(R, \beta_d(R))$) (blue points) is very close to the results obtained at $\beta=\infty$ (pink points).
As R gets closer to  $R_d^\oplus=0.197$, the clustering inverse temperature $\beta_d(R)$ decreases, and the finite-temperature prediction $D_{\min}(R, \beta_d(R))$ increases. 
What is relevant to stress is that BP+decimation is following closely the values of $D_{\min}(R, \beta_d(R))$) and seems to undergo a very similar transition. It looks like if BP were trying to optimize the finite-$\beta$ measure and got stuck at the dynamical transition $\beta_d$.
Obviously when $\beta_d$ is very small (and specially for $R<R_d^\oplus$ when $\beta_d=0$) it is unrealistic than BP can not do better than a random guess, and indeed the distortion reached by BP+decimation is always lower than 0.5 even for small R.

The bottom panels of Fig.~\ref{fig:entropic-dyn-transition} show in more details the RS/RSB transition happening at fixed rate $R=0.25$ (i.e.\ with $\lambda_3=0.25$), for a population of size $10^5$, as $\beta$ increases. 
The overlaps $q_0$ and $q_1$ (see equation (\ref{eq:overlaps}) in appendix~\ref{sec:cavity_method_equations}) are plotted against $\beta$ in the bottom left panel.
One sees that the clustering transition happens at $\beta_d=1.1$: for $\beta<\beta_d$ one has $q_0=q_1$ while for $\beta>\beta_d$ one has $q_1-q_0>0$ strictly.
As soon as we enter in the 1RSB phase, the overlap inside a state $q_1$ becomes equal to $1$. 
This means that the clusters are point-like (the same phenomenon happens for the binary perceptron \cite{krauth1989storage}).

The bottom right panel shows the complexity $\Sigma(x=1)$ (see equation~\ref{eq:complexity_x1}) computed at rate $R=0.25$ and for increasing $\beta$.
The complexity has a large uncertainty and eventually becomes negative for $\beta_c\approx 1.3$, which means that there could be a condensation transition.
From these data it is very hard to say whether there is a RFOT or the transition is eventually FRSB. In any case we have to remind that in models like this one and the binary perceptron where clusters are point-like, the algorithms are unlikely to reach the point-like clusters are much more likely to converge to rare and subdominant clusters of large entropy \cite{braunstein_learning_2006,baldassi2020clustering}

%% file: sections/conclusions.tex
\section{Conclusion}

In this paper we have performed a quantitative study of a random constrained optimization problem called Closest Vector Problem (CVP).
Notwithstanding the simple definition of the model which has linear constraints in $GF(2)$ plus a linear function to optimize, we have uncovered several non trivial phase transitions, that turn out to be related to the performances of the best algorithms searching for optimal solutions (which are message-passing algorithms).

Using the zero-temperature cavity method, we have provided analytical predictions for the minimal energy in several random ensembles of instances. 
For the special case of cycle codes, where variables are involved in at most two constraints, we were able to provide exact results, in particular we proved the exactness of the output of Max-Sum algorithm.

Willing to get closer to the Shannon bound, one has to consider the problem where variables are involved in more than 2 constraints: in this case several interesting phase transition take place lowering the rate.
We have identified two different clustering transitions, one affecting the solution set of the linear system of constraints (occurring at $R_d^\oplus=0.197$), and the other one affecting the set of solutions achieving the minimal energy (occurring at $R_d^{\rm CVP}=1/3$).
More precisely, we found a regime $R\in[R_d^\oplus, R_d^{\rm CVP}]$ in which the solution set of the linear system is rather well connected, yet the energy landscape is glassy and the set optimal solutions is clustered.
Surprisingly, the message-passing algorithms perform well in this regime, and start to be less efficient only in the phase where the solution set of the linear system is itself clustered.
We provide a possible explanation to the above observations based on the different nature of the transition, continuous vs. discontinuous.
It seems that the random first order transition and the strong clustering of solutions happening only for $R<R_d^\oplus$ is at the heart of the computational hardness of the problem.
It is worth studying further how much this is related to the overlap gap property \cite{gamarnik2021overlap}.

%% file: sections/gfq.tex
\section[Moving to GF(q)]{Moving to $\text{GF}(q)$}
\label{sec:gfq}

As shown in the context of channel coding \cite{davey1998low} and source coding \cite{braunstein_efficient_2009}, distortion performance can be improved by working with variables in the finite field $\text{GF}(q)$ with $q=2^k, k\in\mathbb{N}$, as a generalization of binary numbers. 
In this case the parity check matrix $H$ has elements $H_{ai}\in\{0,1,\dots,q-1\}$, and similarly for the source vector $\underline{y}$.

The Boltzmann distribution (\ref{eq:prob_J}) becomes:
\begin{equation}
\mu(\underline{x}) = \frac{1}{Z} \prod_{a=1}^m \mathbb{I}\left[\bigoplus_{j\in\partial a}H_{aj}x_j=0\right] \exp\left( -\beta \sum_{i=1}^n d_H(x_i,s_i)\right)
\end{equation}
where by $\oplus$ we indicate the XOR function, which applies point-wise to the binary representation of $\text{GF}(2^k)$ numbers. The Hamming distance $d_H(x,y)$ between two numbers in $\text{GF}(2^k)$ is the number of ones in the binary representation of $x\oplus y$. 
The operation $H_{aj}x_j$ is intended as the $\text{GF}(2^k)$ multiplication.

The Max-Sum equations in $\text{GF}(q)$ read:
\begin{equation}\label{eq:bp_gfq}
\begin{aligned}
u_{a\to i}(x_i) &=  \max_{\underline{x}_{a\setminus i}:  \bigoplus_{k\in\partial a}H_{ak}x_k  =0}\left[\sum_{j\in\partial a\setminus i} h_{j\to a}(x_j)\right] - \widehat{C}_{a\to i}\\
h_{i\to a}(x_i) &= \sum_{b\in\partial i \setminus a} u_{b\to i}(x_i) - d_H(x_i,y_i) - C_{i\to a}
\end{aligned}
\end{equation}
where $C_{i\to a}$ and $\widehat{C}_{a\to i}$ are constants ensuring respectively that $\max_{x_i}h_{i\to a}(x_i)=0$ and $\max_{x_i}u_{a\to i}(x_i)=0$.

For a source vector $\underline{y}\in\{0,1\}^n$, the corresponding $\text{GF}(2^k)$ counterpart is obtained by grouping bits in groups of $k$ and interpreting them as binary code. 
Of course vector sizes must be adjusted as to produce integer values.
Likewise, given a solution for the problem on $\text{GF}(2^k)$, one unwraps each element into $k$ bits and concatenates them to obtain a binary vector.

\begin{figure}
	\centering
	\includegraphics[width=0.8\columnwidth]{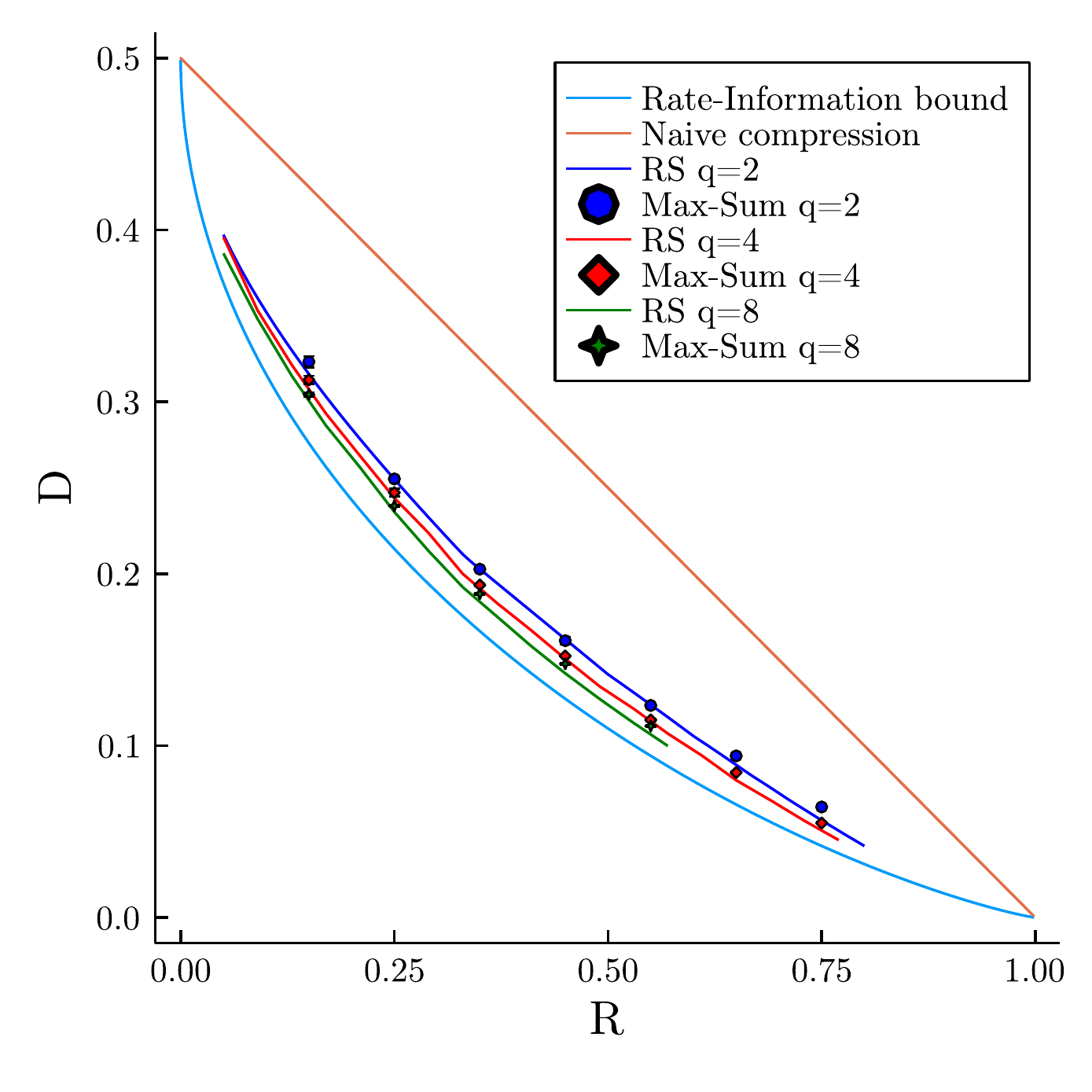}
	\caption{Rate-distortion performance for the $\text{GF}(q)$ Max-Sum algorithm on codes of $n= 6720$ bits (points), plotted against the Replica-Symmetric prediction (plain lines). The degree profile is $\Lambda(x)=\{\lambda_2=1\}$, $P(x)=\{p_k, p_{k+1}\}$. Points are the average over 10 random graphs and source vectors. For each instance, Max-Sum was run from 5 random initializations of messages, the one that gave the minimum number of unsatisfied constraints was kept}
	\label{fig:gfq_MS_RS}
\end{figure}

Figure~\ref{fig:gfq_MS_RS} shows the results of Max-Sum algorithm with reinforcement, plotted against the Replica-Symmetric prediction.
Increasing $q$, one observes indeed that the average minimal distortion decreases.
The results of Max-Sum and the RS prediction starts to differ when $q$ increases, especially at large rate $R$.
This phenomenon was already encountered for $q=2$ (see main text, section~\ref{subsec:deg2_cavityresults} in particular Fig.~\ref{fig:deg2-highrate}.).
We leave for future work the investigation of this discrepancy, as already argued in section~\ref{subsec:deg2_cavityresults} it is possible that the zero-temperature cavity method is not correct in this regime. Another possibility is the apparition of a 1RSB or full RSB phase transition as $q, R$ increases for cycle codes. 

%% file: sections/cavity_method_equations.tex
\section{Equations for the cavity method}
\label{sec:cavity_method_equations}

The aim of the cavity method is to characterize the properties of the probability measure (\ref{eq:prob}), for typical random graphs in $\mathbb{G}_n(\Lambda, P)$ and realization of the external fields $s_1,\dots,s_n$, in the thermodynamic limit.
In this appendix we explain the main steps of the cavity method.
In particular, we describe the zero-temperature cavity method, which focus on the zero-temperature limit at which the probability measure (\ref{eq:prob}) concentrates on the configurations satisfying the constraints and achieving the minimal energy.
We start by describing the finite-temperature version of the cavity method (see \ref{subsec:finite_T_study}).

\subsection{Finite-temperature cavity method}
\label{subsec:entropic_cavity_method}

We start with the study of the properties of (\ref{eq:prob}) at finite $\beta$. 
The cavity method allows to compute the quenched free-energy density:
\begin{align}
	F(\Lambda, P, \beta) = \lim_{n\to\infty}\frac{1}{n}\mathbb{E}_{\cG, \underline{s}}[F(\cG, \underline{s}, \beta)]
\end{align}
and the average energy:
\begin{align}
	E(\Lambda, P, \beta) = \lim_{n\to\infty}\frac{1}{n}\mathbb{E}_{\cG, \underline{s}}[\langle E(\sigma)\rangle]
\end{align}
where $\mathbb{E}_{\cG, \underline{s}}$ is the average over the random ensemble of XORSAT instances defined by a bipartite graph $\cG$ drawn from $\mathbb{G}_n(\Lambda, P)$ and a set of external fields $\underline{s}$ drawn independently and uniformly in $\{-1,1\}$, and the bracket is the average over the measure $\mu$ (defined in \ref{eq:prob}).

\subsubsection{Replica Symmetric cavity method}

There are several version of the cavity method. 
In the simplest version, called Replica-Symmetric (RS), we assume a fast decay of the correlations between distant variables in the measure $\mu(\us)$ defined in (\ref{eq:prob}) in such a way that the BP equations converge to a unique fixed-point on a typical large instance.
Consider an uniformly chosen directed edge $i\to a$ in a random bipartite graph $\cG$, and call $P^{\rm rs}$ the probability law of the fixed-point message $m_{i\to a}$ thus obtained. Similarly, call $Q^{\rm rs}$ the probability of the message $\hm_{a\to i}$. Within the decorrelation hypothesis of the RS cavity method, the incoming messages on a given variable node (resp. factor node) are i.i.d. with the law $Q^{\rm rs}$ (resp. $P^{\rm rs}$). This implies that the laws $P^{\rm rs}$ and $Q^{\rm rs}$ must obey the following equations:
\begin{align}\label{eq:RS_entropic}
\begin{aligned}
    P^{\rm rs}(m) &= \sum_{d=0}^{\infty}\tl_d\sum_{s}\frac{1}{2}\int\prod_{a=1}^d\dd Q^{\rm rs}(\hm_a) \\
    &\times\delta(m-f^{\rm bp}(\hm_1,\dots,\hm_d;s))\\
    Q^{\rm rs}(\hm) &= \sum_{k=0}^{\infty}\tp_k\int\prod_{i=1}^k\dd P^{\rm rs}(m_i) \delta(\hm-\hat{f}^{\rm bp}(m_1,\dots,m_k))
\end{aligned}
\end{align}
where $f^{\rm bp}(\hm_1,\dots,\hm_d;s)$ and $\hat{f}^{\rm bp}(m_1,\dots,m_k)$ are shorthand notations for the r.h.s of the BP equations (\ref{eq:BP}), and $\tl_d, \tp_k$ are the residual degrees:
\begin{align}
	\tl_d = \frac{d\lambda_d}{\sum_{i=1}^{d_{\max}}i\lambda_i} \ , \quad \tp_k = \frac{kp_k}{\sum_{i=1}^{k_{\max}}ip_i} 
\end{align}
We numerically solved these equations with population dynamics. 
The RS cavity prediction of the quenched free-energy $F(\Lambda, P, \beta)$ is then obtained by averaging the Bethe expression (\ref{eq:BetheF}) with respect to the message distributions $P^{\rm rs}$ and $Q^{\rm rs}$:
\begin{align}\label{eq:RS_freeen}
\begin{aligned}
    -\beta &F^{\rm rs}(\Lambda, P, \beta) = \sum_{s}\frac{1}{2}\sum_{d=1}^{d_{\max}}\lambda_d\int\prod_{a=1}^d\dd Q^{\rm rs}(\hm_a)\\
    &\times\log Z_i(\hm_1,\dots,\hm_d;s)\\
    &+\alpha\sum_{k=1}^{k_{\max}}p_k\int\prod_{i=1}^k\dd P^{\rm rs}(m_i)\log Z_a(m_1,\dots,m_k) \\
    &-d_{\rm avg}\int\dd P^{\rm rs}(m)\dd Q^{\rm rs}(\hm)\log Z_{ia}(m,\hm)
\end{aligned}
\end{align}
with $\alpha=\frac{\sum_{i=1}^{d_{\max}}i\lambda_i}{\sum_{i=1}^{k_{\max}}ip_i}$ the density of constraints, $d_{\rm avg}=\sum_{i=1}^{d_{\max}}i\lambda_i$ the mean variable degree, and $Z_i, Z_a, Z_{ia}$ defined in (\ref{eq:BetheF_defterms})
Similarly, the RS prediction of the average internal energy is:
\begin{align}
\label{eq:int_nrj_RSentropic}
\begin{aligned}
E^{\rm rs}(\Lambda, P, \beta) =& -\sum_{s}\frac{1}{2}\sum_{d=1}^{d_{\max}}\lambda_d\int\prod_{a=1}^d\dd Q^{\rm rs}(\hm_a)\\
	&\times\frac{\sum_{\sigma}s\sigma e^{\beta\sigma s}\prod_{a=1}^d\hm_a(\sigma)}{\sum_{\sigma}e^{\beta\sigma s}\prod_{a=1}^d\hm_a(\sigma)}
\end{aligned}	
\end{align}

\subsubsection{1RSB formalism}

The hypothesis underlying the RS cavity method must break down when the density of constraints $\alpha$ become too large, leading to the Replica Symmetry Breaking (RSB) phenomenon. 
In such a case it becomes necessary to use more refined versions of the cavity method to describe the typical properties of the problem. 
The first non-trivial level is called the one step Replica-Symmetry Breaking (1RSB) cavity method. 
It postulates the decomposition of the configuration space into clusters, such that the restriction of the measure $\mu$ to a cluster has good decorrelation properties and can be described by the RS cavity method. 
More precisely, we index by $\gamma$ the partition of the configuration space into clusters, and denote $Z_{\gamma}(\beta)$ the contribution to the partition function $Z(\beta)$ of the $\gamma$-th cluster, as well as $m_{i\to a}^{\gamma},\hm_{a\to i}^{\gamma}$ the solution of the BP equations that describe it. 
The 1RSB cavity method aims at computing the potential
\begin{align}
\label{eq:1RSBpotential}
    \Phi^{\rm 1rsb}(x, \beta)=\lim_{n\to\infty}\frac{1}{n}\mathbb{E}_{\cG, \underline{s}}\left[\ln\left(\sum_{\gamma}(Z_{\gamma}(\beta))^x\right)\right]
\end{align}
The Parisi parameter $x$ allows to weight differently the various clusters. 
The original problem is described by the choice $x=1$. 
The potential (\ref{eq:1RSBpotential}) contains precious information about the cluster decomposition.
Suppose that, at the leading exponential order, there are $\Sigma(\phi)$ clusters $\gamma$ with $Z_{\gamma}=e^{N\phi}$ (neglecting sub-exponential corrections). 
The complexity $\Sigma(\phi)$ plays the role of the entropy density, in an auxiliary model where the clusters are replacing usual configurations. 
The potential $\Phi^{\rm 1rsb}(x)$ and the complexity $\Sigma(\phi)$ are Legendre transform of each other:
\begin{align}
	\label{eq:1rsb_potential}
    \Phi^{\rm 1rsb}(x)=\sup_{\phi}[\Sigma(\phi)+x\phi]
\end{align}
This relation can be inverted:
\begin{align}
	\Sigma(x) = \Phi^{\rm 1rsb}(x) -x\frac{\dd}{\dd m}\Phi^{\rm 1rsb}(x)
\end{align}
One introduces, for a given sample and a given edge $(i,a)$ of the bipartite graph $\cG$, two distributions $P_{i\to a}$ and $Q_{a\to i}$ that encode the laws $m_{i\to a}^{\gamma}$ and $\hm_{a\to i}^{\gamma}$ when the cluster $\gamma$ is chosen randomly with probability proportional to $(Z_{\gamma})^x$. These distributions obey the self-consistent equations:
\begin{align}\label{eq:1RSB_xgen}
\begin{aligned}
    &P_{i\to a}(m_{i\to a}) = \frac{1}{Z^{\rm 1rsb}_{i\to a}}\int\prod_{b\in\dima}\dd Q_{b\to i}(\hm_{b\to i}) \\
    \times &\delta[m_{i\to a}-f^{\rm bp}(\{\hm_{b\to i}\}_{b\in\dima};s_i)]z_{i\to a}(\{\hm_{b\to i}\}_{b\in\dima};s_i)^x\\
    &Q_{a\to i}(\hm_{a\to i}) = \frac{1}{\hat{Z}^{\rm 1rsb}_{a\to i}}\int\prod_{j\in\dami}\dd P_{j\to a}(m_{j\to a}) \\
    \times &\delta[\hm_{a\to i}-\hat{f}^{\rm bp}(\{m_{j\to a}\}_{j\in\dami})]\hat{z}_{a\to i}(\{m_{j\to a}\}_{j\in\dami})^x
\end{aligned}
\end{align}
where $z_{i\to a}$, $\hat{z}_{a\to i}$ are the normalization constants in the BP equations (\ref{eq:BP}), and the factors $Z^{\rm 1rsb}_{i\to a}$, $\hat{Z}^{\rm 1rsb}_{a\to i}$ ensure the normalization of $P_{i\to a}$ and $Q_{a\to i}$ respectively.
In order to average over the random ensemble of instances one introduces the probability distributions over the 1RSB messages $\cP^{\rm 1rsb}(P)$ and $\cQ^{\rm 1rsb}(Q)$, that obey the consistency relations similar to (\ref{eq:RS_entropic}):
\begin{align}
\label{eq:1rsb_entropic_x}
\begin{aligned}
    \cP^{\rm 1rsb}(P)&=\sum_{d=0}^{\infty}\tl_d\sum_s\frac{1}{2}\int\prod_{a=1}^d\dd \cQ^{1rsb}(Q_a)\\
    &\times\delta[P-F(Q_1,\dots,Q_d;s)]\\
    \cQ^{\rm 1rsb}(Q)&=\sum_{k=0}^{\infty}\tp_k\int\prod_{i=1}^k\dd \cP^{\rm 1rsb}(P_i)\delta[Q-\hat{F}(P_1,\dots,P_k)]
\end{aligned}
\end{align}
where $F(Q_1,\dots,Q_d;s)$ (resp. $\hat{F}(P_1,\dots,P_k)$) is a shorthand notation for the r.h.s. of the first (resp. second) equation in (\ref{eq:1RSB_xgen}). 
The 1RSB potential (\ref{eq:1rsb_potential}) can be computed from the solution of these equations:
\begin{align}
\begin{aligned}
	&\Phi^{\rm 1rsb}(x, \beta) = \alpha \sum_{k=1}^{k_{\max}}\int \prod_{i=1}^k\dd\cP^{\rm 1rsb}(P_i)\log Z^{\rm 1rsb}_a(P_1, \dots, P_k) \\
	&+ \sum_s\frac{1}{2}\sum_{d=1}^{d_{\max}}\lambda_d\int\prod_{a=1}^d\dd\cQ^{\rm 1rsb}(Q_a) \log Z^{\rm 1rsb}_i(Q_1, \dots, Q_d; s)\\
	&-d_{\rm avg}\int\dd\cP^{\rm 1rsb}(P)\dd\cQ^{\rm 1rsb}(Q)\log Z_{ia}^{\rm 1rsb}(P, Q)
\end{aligned}
\end{align}
with
\begin{align}
\begin{aligned}
	Z^{\rm 1rsb}_a(P_1, \dots, P_k) &= \int\prod_{i=1}^k\dd P_a(m_a)Z_a(m_1, \dots, m_k)^x\\
	Z^{\rm 1rsb}_i(Q_1, \dots, Q_d; s) &= \int\prod_{a=1}^d\dd Q_i(\hm_i)Z_i(\hm_1, \dots, \hm_d; s)^x\\
	Z_{ia}^{\rm 1rsb}(P, Q) &=\int\dd P(m)\dd Q(\hm) Z_{ia}(m, \hm)^x
\end{aligned}
\end{align}
with $Z_i, Z_a, Z_{ia}$ defined in (\ref{eq:BetheF_defterms}).
Finally the 1RSB prediction of the quenched free-entropy is:
\begin{align}
\label{eq:extremum}
	-\beta F^{\rm 1rsb} = \underset{x\in[0,1]}{\inf}\left[\frac{\Phi^{\rm 1rsb}(x)}{x}\right]
\end{align}
Note that the 1RSB equations (\ref{eq:1rsb_entropic_x}) always admit the RS solution as a trivial fixed point, in which the distributions $P$ (resp. $Q$) in the support of $\cP^{\rm 1rsb}$ (resp. $\cQ^{\rm 1rsb}$) are Dirac measures:
\begin{align}
	\label{eq:trivial_RS}
	\cP^{\rm 1rsb}(P) = \int \dd P^{\rm rs}(h)\delta[P-\delta(\cdot - h)]
\end{align}
with $P^{\rm rs}$ solution of the RS equation (\ref{eq:RS_entropic}) and similarly for $\cQ^{\rm 1rsb}$.
For small values of the density of constraints $\alpha$, this trivial solution is the only one, then $F^{\rm 1rsb}=F^{\rm rs}$ and the predictions given by the RS and 1RSB cavity method coincide: we are in the RS phase.
Increasing $\alpha$, non-trivial solutions of (\ref{eq:1rsb_entropic_x}) can appear. The clustering threshold $\alpha_d(\beta)$ is defined as the smallest value of $\alpha$ for which the 1RSB equations at $x=1$ admit a non-trivial solution.
If the associated complexity $\Sigma(x=1)$ is positive, then the extremum in (\ref{eq:extremum}) is reached for $x=1$ and therefore $F^{\rm 1rsb} = F^{\rm rs}$. 
In that case the typical solutions of the measure (\ref{eq:prob}) have split into an exponential number of clusters, in such a way that the total free-energy is unable to detect the difference with the RS situation.
This phase is called the dynamic 1RSB phase.
If instead one has $\Sigma(x=1)<0$, then the extremum in (\ref{eq:extremum}) is is reached at a non-trivial value $x_s<1$ of the Parisi parameter. The number of clusters is sub-exponential, and the computation of the free-energy allows to detect the 1RSB phenomenon. 
One calls condensation threshold $\alpha_c(\beta)$ the smallest value of $\alpha$ for which a solution of the 1RSB equations (\ref{eq:1rsb_entropic_x}) with $\Sigma(x=1)<0$ exists.

\subsubsection{Simplifications for $x=1$}
As explained in the previous sub-section the dynamical-1RSB phase is well described by the choice $x=1$ for the Parisi parameter.
We have focused on this paper on the computation of the clustering threshold $\alpha_d(\beta)$, which allows us to restrict our analysis to $x=1$.
We can also detect the condensation phenomenon by computing the complexity at $x=1$ and looking at the first value $\alpha_c$ for which it becomes negative.
The complete 1RSB equations can be largely simplified for the special value of the parameter $x=1$. As explained in \cite{montanari2008clusters}, the first step is to note that the normalizations $Z^{\rm 1rsb}_{i\to a}$ (resp. $\hat{Z}^{\rm 1rsb}_{a\to i}$) depends only on the distributions $\{Q_{b\to i}\}_{b\in\dima}$ (resp. on $\{P_{j\to a}\}_{j\in\dami}$) through their mean value. Define $\lm[P], \lhm[Q]$ as the averages:
\begin{align}
	\lm[P](\sigma)=\int\dd P(m) m(\sigma), \ \lhm[Q](\sigma)=\int\dd Q(\hm)\hm(\sigma)
\end{align}
then one can check that the normalization constant $Z^{\rm 1rsb}_{i\to a}$ depends on the distributions $Q_1,\dots, Q_d$ only through the averages $\lhm[Q_1],\dots,\lhm[Q_d]$ (and similarly for $\hat{Z}^{\rm 1rsb}_{a\to i}$)
\begin{align}
\begin{aligned}
	Z^{\rm 1rsb}_{i\to a}(Q_1,\dots,Q_d,s) &= z_{i\to a}(\lhm[Q_1],\dots,\lhm[Q_d],s)\\
    \hat{Z}^{\rm 1rsb}_{a\to i}(P_1,\dots,P_k) &= \hat{z}_{a\to i}(\lm[P_1],\dots,\lm[P_k])=1
\end{aligned}
\end{align}
Then, one can check that the random variable $\lm[P]$, $\lhm[Q]$ obtained by drawing $P$ (resp. $Q$) from $\cP^{\rm 1rsb}(P)$ (resp. $\cQ^{\rm 1rsb}(Q)$) obey the RS equations (\ref{eq:RS_entropic}), and therefore are distributed according to the RS distributions $P^{\rm rs}, Q^{\rm rs}$. 
One defines the conditional averages:
\begin{align}
\begin{aligned}
	P(m|\lm) = \frac{1}{P^{\rm rs}(\lm)}\int \dd \cP^{\rm 1rsb}(P) P(m)\delta(\lm-\lm[P]) \\
	Q(\hm|\lhm) = \frac{1}{Q^{\rm rs}(\lhm)}\int \dd \cQ^{\rm 1rsb}(Q) Q(\hm)\delta(\lhm-\lhm[Q])
\end{aligned}
\end{align}
We can get closed equations on these distributions (that we will not write here, see \cite{montanari2008clusters} for a complete derivation of these equations), that still have a reweighting term. To get rid of it one defines
\begin{align}
\begin{aligned}
	P_{\sigma}(m|\lm) &= \frac{m(\sigma)}{\lm(\sigma)}P(m|\lm), \\
    Q_{\sigma}(\hm|\lhm) &= \frac{\hm(\sigma)}{\lhm(\sigma)}Q(\hm|\lhm)
\end{aligned}
\end{align}
and obtains the following closed equations:
\begin{widetext}
\begin{align}
\label{eq:1rsb_entropic_x1}
\begin{aligned}
    P_{\sigma}(m|\bar{m})P^{\rm rs}(\bar{m}) &= \sum_{d=0}^{d_{\max}-1}\tl_d\sum_{s}\frac{1}{2}\int\prod_{a=1}^d\dd Q^{\rm rs}(\bar{\hm}_a)\delta(\bar{m}-f^{\rm bp}(\bar{\hm}_1,\dots,\bar{\hm}_d;s)\int\prod_{a=1}^d\dd Q_{\sigma}(\hm_a|\bar{\hm}_a)\\
    &\times\delta(m-f^{\rm bp}(\hm_1,\dots,\hm_d;s) \\
    Q_{\sigma}(\hm|\bar{\hm})Q^{\rm rs}(\bar{\hm}) &= \sum_{k=1}^{k_{\max}-1}\tp_k\int\prod_{i=1}^k\delta(\bar{\hm}-\hat{f}^{\rm bp}(\bar{m}_1,\dots\bar{m}_k))\sum_{\sigma_1,\dots,\sigma_d}\nu(\sigma_1,\dots,\sigma_d|\bar{m}_1,\dots,\bar{m}_k,\sigma)\\
    &\times \int\prod_{i=1}^k\dd P(m_i|\bar{m}_i)\delta(\hm-\hat{f}^{\rm bp}(m_1,\dots m_k))
\end{aligned}
\end{align}
where
\begin{align}
    \nu(\sigma_1,\dots,\sigma_d|\bar{m}_1,\dots,\bar{m}_k,\sigma) = \frac{\mathbb{I}\left[\sigma\prod_{i=1}^k\sigma_i=1\right]\prod_{i=1}^k\bar{m}_i(\sigma_i)}{\sum_{\sigma_1,\dots,\sigma_k}\mathbb{I}\left[\sigma\prod_{i=1}^k\sigma_i=1\right]\prod_{i=1}^k\bar{m}_i(\sigma_i)}
\end{align}
\end{widetext}
The above equations can be solved with population dynamics, with two populations of triples $\{(\lm_i, m_i^+, m_i^-): i=1,\dots, \mathcal{N}\}$  and $\{(\lhm_i, \hm_i^+, \hm_i^-): i=1,\dots, \mathcal{N}\}$. 
In this formalism, the trivial RS solution (\ref{eq:trivial_RS}) of the 1RSB equations takes the following form:
\begin{align}
\begin{aligned}
	P_{\sigma}(m|\bar{m})P^{\rm rs}(\bar{m}) &= \delta(m-\bar{m})P^{\rm rs}(\bar{m}) \\
	Q_{\sigma}(\hm|\bar{\hm})Q^{\rm rs}(\bar{\hm})& = \delta(\hm-\bar{\hm})Q^{\rm rs}(\bar{\hm})
\end{aligned}
\end{align}
We use the intra-state overlap $q_0$ and the inter-state overlap $q_1$ to determine the apparition of the dynamical-1RSB phase at $\alpha_d(\beta)$:
\begin{align}
\label{eq:overlaps}
\begin{aligned}
	q_0 &=\int\dd P^{\rm rs}(\bar{m})\left(\sum_{\sigma}\sigma \bar{m}(\sigma)\right)^2 \\ 
	q_1 &=\int\dd P^{\rm rs}(\bar{m})\sum_{\sigma}\sigma\bar{m}(\sigma)\int\dd P_{\sigma}(m|\bar{m})\left(\sum_{\sigma'}\sigma'm(\sigma')\right)
\end{aligned}
\end{align}
In the RS phase one has $q_0=q_1$, while in the dynamical-1RSB phase $q_0>q_1$ strictly.
One can compute the thermodynamic quantities from the solution of equation (\ref{eq:1rsb_entropic_x1}).
At $x=1$, one can check that the 1RSB potential (\ref{eq:1RSBpotential}) equals the RS free-entropy: $\Phi^{\rm 1rsb}(x=1)=\phi^{\rm rs}=-\beta F^{\rm rs}(\beta)$ (with $F^{\rm rs}(\beta)$ defined in equation (\ref{eq:RS_freeen})). 
The 1RSB internal free energy, i.e. the average over the clusters of the Free Energy associated to one cluster can be computed from the 1RSB solution at $x=1$ (see equation (54) of \cite{montanari2008clusters}):
\begin{widetext}
\begin{align}
\begin{aligned}
    \phi_{\rm int}(x=1) &=\sum_{d=1}^{d_{\max}}\lambda_d \sum_s\frac{1}{2}\int \prod_{a=1}^d\dd Q^{\rm rs}(\bar{\hm}_a)\sum_{\sigma}\frac{e^{\beta s\sigma}\prod_{a=1}^k\lhm_a(\sigma)}{Z_i(\lhm_1,\dots,\lhm_d,s)}\int \prod_{a=1}^d\dd Q_{\sigma}(\hm_a|\lhm_a)\log(Z_i(\hm_1,\dots,\hm_d,s)) \\
    + \alpha\sum_{k=1}^{k_{\max}} &p_k\int \prod_{i=1}^k\dd P^{\rm rs}(\lm_i)\sum_{\sigma_1,\dots,\sigma_k}\frac{\mathbb{I}[\prod_{i=1}^k\sigma_i=1]\prod_{i=1}^k\lm_i(\sigma_i)}{Z_a(\lm_1,\dots,\lm_k)}\int \prod_{i=1}^k\dd P_{\sigma_i}(m_i|\lm_i)\log(Z_a(m_1,\dots,m_k)) \\
    &- d_{\rm avg} \int\dd P^{\rm rs}(\lm)\dd Q^{\rm rs}(\lhm)\sum_{\sigma}\frac{\lm(\sigma)\lhm(\sigma)}{Z_{ia}(\lm,\lhm)}\int\dd Q_{\sigma}(\hm|\lhm)\dd P_{\sigma}(m|\lm)\log Z_{ia}(m,\hm)
\end{aligned}
\end{align}
\end{widetext}
From the 1rsb potential $\Phi^{\rm 1rsb}$ and the internal free energy $\phi_{\rm int}$ one can deduce the complexity
\begin{align}
    \label{eq:complexity_x1}
    \Sigma(x=1) = \Phi^{\rm 1rsb}(x=1) - \phi_{\rm int}(x=1)
\end{align}
Finally, the 1RSB prediction for the internal distortion, i.e. the average over the clusters of the distortion associated to one cluster can also be computed at $x=1$. One can check that it equals the RS prediction of the average distortion $D_{\rm int}(x=1) = D^{\rm rs}=(1+E^{\rm rs})/2$ with, with $E^{\rm rs}$ defined in equation (\ref{eq:int_nrj_RSentropic}).

\subsection{Zero-temperature cavity method}
\label{subsec:en_cavity_method}
We are interested in the computation of the average (over the random ensemble of instances) of the averaged energy $\langle E(\us)\rangle$ over the measure (\ref{eq:prob}) in the large $\beta$ limit:
\begin{align}
	E^{\min} = \lim_{n\to\infty}\frac{1}{n}\mathbb{E}_{\cG, \underline{s}}\left[\lim_{\beta\to\infty}\langle E(\us)\rangle_{\mu_{\beta}}\right]
\end{align}
In this limit one obtains the minimal energy, or equivalently the minimal distance $D^{\min}=(1+E^{\min})/2$ achieved between the source and the closest codeword, in average over the random ensemble of instances.
A simplified version of the cavity method can be developed in this limit, called the zero-temperature cavity method.
We will work with the softened version of the probability measure $\mu_J$ defined in (\ref{eq:prob_J}), with $J$ a real parameter that is conjugated to the number of unsatisfied constraints, because it allows to simplify the numerical resolution of the zero-temperature RS and 1RSB equations.

\subsubsection{Replica-Symmetric formalism}
In the RS formalism, one assumes that the Max-Sum equations (\ref{eq:MS}, \ref{eq:MS_J}) admit a unique solution, that describe correctly the properties of the measure (\ref{eq:prob_J}) in the large $\beta$ limit. 
For a uniformly chosen directed edge $i\to a$ in a random bipartite graph $\cG$, we define $p^{\rm rs}$ the probability law of the fixed-point message $h_{i\to a}$, and similarly the probability law $q^{\rm rs}$ of the message $u_{a\to i}$. The laws $p^{\rm rs}$ and $q^{\rm rs}$ obey the following self-consistency relations:
\begin{align}\label{eq:RS_energetic}
\begin{aligned}
p^{\rm rs}(h) &= \sum_{d=0}^{\infty}\tl_d\sum_{s}\frac{1}{2}\sum_{u_1,\dots,u_d} \delta(h-f^{\rm ms}(u_1,\dots,u_d;s))\\
&\times\prod_{a=1}^d q^{\rm rs}(u_a)\\
q^{\rm rs}(u) &= \sum_{k=0}^{\infty}\tp_k\sum_{h_1,\dots,h_k} \delta(u-\hat{f}^{\rm ms}(h_1,\dots,h_k)) \prod_{i=1}^k p^{\rm rs}(h_i)
\end{aligned}
\end{align}
where $f^{\rm ms}(u_1,\dots,u_d;s)$ and $\hat{f}^{\rm ms}(h_1,\dots,h_k)$ are shorthand notation for the r.h.s. of the MS equations (\ref{eq:MS} and \ref{eq:MS_J}).
The above equations can be re-written in a simplified form. Define $p_0(h) =\frac{1}{2}(\delta(h,1)+\delta(h,-1))$ and let $\circledast$ be the convolution operation. We obtain for $p^{\rm rs}(h)$:
\begin{equation}
\label{eq:RS_energetic_p}
p^{\rm rs} = \sum_{d=1}^{d_{\max}}\tl_d \left(p_0\circledast \underbrace{q^{\rm rs} \circledast\dots\circledast q^{\rm rs}}_{d \text{ times}}\right)
\end{equation}
while we obtain for $q^{\rm rs}(u)$, when $u>0$:
\begin{equation}
\label{eq:RS_energetic_q}
\begin{aligned}
q^{\rm rs}(u) &= \sum_{k=1}^{k_{\max}}\tp_k 2^{k-1}\left\{\left[\sum_{h\ge u}p^{\rm rs}(h)\right]^k -\left[\sum_{h> u}p^{\rm rs}(h)\right]^k \right\} \\
{\rm and} \ &q^{\rm rs}(0) = \sum_{k=1}^{k_{\max}}\tp_k\left(1 - \left[1-p^{\rm rs}(0)\right]^{k}\right) \ {\rm when} \ u=0 \ .
\end{aligned}
\end{equation}
With a finite parameter $J$, the factor-to-variable $u_{a\to i}$ take integer values in $[-J,J]$ while the variable-to-factor messages $h_{i\to a}$ take integer values in $[-1-J(d_{\max}-1),1+J(d_{\max}+1)]$. 
The equations (\ref{eq:RS_energetic_p}, \ref{eq:RS_energetic_q}) can be solved numerically in an iterative way, with real vectors representing the distributions $p^{\rm rs}$ and $q^{\rm rs}$. 
Once a solution to the RS equations is found, the RS cavity prediction of the minimal Energy $E^{\min}$ is then obtained by averaging the Bethe expression with respect to the message distributions $p^{\rm rs}$ and $q^{\rm rs}$:
\begin{align}
E^{\rm rs} &= \sum_{d=1}^{d_{\max}}\lambda_k\sum_{u_1,\dots,u_k}E_i(u_1,\dots,u_d)\prod_{a=1}^d q^{\rm rs}(u_a)\\
&+ \alpha\sum_{k=0}^{\infty}p_k\sum_{h_1,\dots,h_k}E_a(h_1,\dots,h_k) \prod_{i=1}^k p(^{\rm rs}h_i)\\
&- d_{\rm avg}\sum_{h,u}E_{ia}(h, u) p^{\rm rs}(h)q^{\rm rs}(u)    
\end{align}
with $E_i, E_a, E_{ia}$ defined in (\ref{eq:EBethe_defterms}, \ref{eq:EBethe_defterms_J}).

\subsubsection{1RSB formalism}

In the 1RSB zero-temperature formalism one takes simultaneously the limit $x\to 0$ and $\beta\to\infty$ with a fixed finite value of a new parameter $y=\beta x$.
Taking this limit in the expression of the 1RSB potential $\Phi^{\rm 1rsb}(x)$ (equation (\ref{eq:1rsb_potential})) one obtains the zero-temperature version of the 1RSB potential:
\begin{align}
\label{eq:1rsb_energetic_potential}
\Phi_e^{\rm 1rsb}(y) = -yF_e^{\rm 1rsb}(y) = \sup_E\left[\Sigma_e(E)-yE\right]
\end{align}
The zero-temperature complexity $\Sigma_e(E)$ counts the (exponential) number of clusters with a given minimal energy $E$, these clusters being counted independently of their size.
It can be computed via an inverse Legendre transform of the potential $\Phi_e(y)$:
\begin{align}
\Sigma_e(y) = \Phi_e^{\rm 1rsb}(y) - y\frac{\dd}{\dd y}\Phi_e^{\rm 1rsb}(y)
\end{align}
As in the finite $\beta$ formalism, we assume the partition of the configuration space into clusters indexed by $\gamma$, in such a way that each cluster is described by a solution of the Max-Sum equations (\ref{eq:MS}, \ref{eq:MS_J}).
The parameter $y$ allows to weight the clusters according to their Bethe Energy $E_{\gamma} = E^{\rm Bethe}(\{h_{i\to a}^{\gamma}, u_{a\to i}^{\gamma}\})$ (see equation (\ref{eq:EBethe})).
We introduce for a given sample and a given edge $(i,a)$ of the factor graph, two distributions $p_{i\to a}^{\gamma}$ and $q_{a\to i}^{\gamma}$ that encode the laws $h_{i\to a}^{\gamma}$ and $u_{a\to i}^{\gamma}$ when the cluster $\gamma$ is chosen randomly with probability proportional to $e^{-yE_{\gamma}}$. These distributions obey the following self-consistent equations (similar to the equations for the finite-temperature case (\ref{eq:1RSB_xgen})):
\begin{align}
\label{eq:1rsb_energetic_singleinstance}
\begin{aligned}
    &p_{i\to a}(h_{i\to a}) = \frac{e^{y|h_{i\to a}|}}{z^{\rm 1rsb}_{i\to a}}\sum_{\{u_{b\to i}\}_{b\in\dima}}\prod_{b\in\dima}q_{b\to i}(u_{b\to i})\\
        &\times \left(\prod_{b\in\dima}e^{-y|u_{b\to i}|}\right)\delta(h_{i\to a}-f^{\rm ms}(\{u_{b\to i}\}_{b\in\dima};s_i))\\
    &q_{a\to i}(u_{a\to i}) =\frac{1}{\hat{z}^{\rm 1rsb}_{a\to i}}\sum_{\{h_{j\to a}\}_{j\in\dami}}\prod_{j\in\dami}p_{j\to a}(h_{j\to a})\\
    &\times\delta(u_{a\to i}-\hat{f}^{\rm ms}(\{h_{j\to a}\}_{j\in\dami}))
\end{aligned}
\end{align}
We can now average over the random ensemble of instances. Let $(i,a)$ be a uniformly chosen edge in a factor graph drawn from the random ensemble with degree profile $(\Lambda, P)$. One introduces the probability distributions over the 1RSB messages $\cP_e^{\rm 1rsb}(p)$ and $\cQ_e^{\rm 1rsb}(q)$ over the messages $p_{i\to a}, q_{a\to i}$. They obey the consistency relations, that are the zero-temperature version of the 1RSB equations (\ref{eq:1rsb_entropic_x}):
\begin{align}
\label{eq:1RSB_energetic}
\begin{aligned}
    \cP_e^{\rm 1rsb}(p) &= \sum_{d=1}^{\infty}\tl_d\sum_s\frac{1}{2}\int\prod_{b=1}^d\dd \cQ_e^{\rm 1rsb}(q_b)\\
    &\times \delta[p-F_e(q_1,\dots,q_d;s)]\\
    \cQ_e^{\rm 1rsb}(q) &= \sum_{k=1}^{\infty}\tp_k\int\prod_{j=1}^k\dd\cP_e^{\rm 1rsb}(p_j)\delta[q-\hat{F}_e(p_1,\dots,p_k)]
\end{aligned}
\end{align}
where $F_e(q_1,\dots,q_d;s)$ (resp. $\hat{F}_e(p_1,\dots,p_k)$) is a shorthand notation for the r.h.s. of the first (resp. the second) equation in (\ref{eq:1rsb_energetic_singleinstance}).
These equations always admits a trivial fixed-point 
\begin{align}
\label{eq:trivial_RS_energetic}
\begin{aligned}
	\cP_e^{\rm 1rsb}(p) = \sum_h p^{\rm rs}(h)\delta[p-\delta(\cdot-h)] \\
	\cQ_e^{\rm 1rsb}(q) = \sum_h q^{\rm rs}(u)\delta[q-\delta(\cdot-u)] \ ,
\end{aligned}
\end{align} with $p^{\rm rs}$ and $q^{\rm rs}$ solution of the RS equation (\ref{eq:RS_energetic}). 
In the RS phase, this trivial fixed-point is the unique solution, while in the 1RSB phase, the trivial solution becomes unstable and the above equations admits a non-trivial solution.
The clustering transition in the $\beta\to\infty$ limit is therefore defined as the smallest density of constraints $\alpha$ such that the 1RSB equations (\ref{eq:1RSB_energetic}) admit a non-trivial solution.
The 1RSB potential $\Phi_e^{\rm 1rsb}(y)$ can be computed from the solution of these equations:
\begin{align}
\begin{aligned}
    &\Phi_e^{\rm 1rsb}(y) = - d_{\rm avg}\int\dd\cP_e^{\rm 1rsb}(p)\cQ_e^{\rm 1rsb}(q) \phi_{e,ia}(p,q;y) \\ &+\sum_{d=0}^{\infty}\lambda_d\sum_s\frac{1}{2}\int\prod_{a=1}^d\dd\cQ_e^{\rm 1rsb}(q_a)\phi_{e,i}(q_1,\dots,q_d;s,y) \\
    &+ \alpha\sum_{k=0}^{\infty}p_k\int\prod_{i=1}^k\dd\cP_e^{\rm 1rsb}(p_i)\phi_{e,a}(p_1,\dots,p_k;y) 
\end{aligned}
\end{align}
with:
\begin{align}
\begin{aligned}
	\phi_{e,i} &= \log\left(\sum_{u_1,\dots,u_d}e^{-yE_i(u_1,\dots,u_d;s)}\prod_{i=1}^dq_i(u_i)\right)\\
	\phi_{e,a} &= \log\left(\sum_{h_1,\dots,h_k}e^{-yE_a(h_1,\dots,h_k)}\prod_{a=1}^kp_a(h_a)\right)\\
	\phi_{e,ia} &= \log\left(\sum_{h,u}e^{yE_{ia}(h,u)}p(h)q(u)\right)
\end{aligned}
\end{align}
with $E_i, E_a, E_{ia}$ the terms in the Bethe Energy defined in (\ref{eq:EBethe_defterms}, \ref{eq:EBethe_defterms_J}).
One can also compute the internal energy, which is the average (over the clusters) of the Bethe Energy associated to one cluster:
\begin{align}
\label{eq:internal_energy}
\begin{aligned}
    &U_{\rm int}^{\rm 1rsb}(y) = - d_{\rm avg}\int\dd\cP_e^{\rm 1rsb}(p)\cQ_e^{\rm 1rsb}(q) U_{ia}(p,q;y) \\ &+ \sum_{d=1}^{d_{\max}}\lambda_d\sum_s\frac{1}{2}\int\prod_{a=1}^d\dd\cQ_e^{\rm 1rsb}(q_a)U_i(q_1,\dots,q_d;s,y) \\
    &+ \alpha\sum_{k=0}^{\infty}p_k\int\prod_{i=1}^k\dd\cP_e^{\rm 1rsb}(p_i)U_a(p_1,\dots,p_k;y) 
\end{aligned}
\end{align}
with
\begin{align}
\begin{aligned}
	U_i &= \sum_{u_1,\dots,u_d}\frac{E_i(u_1,\dots,u_d;s)e^{-yE_i(u_1,\dots,u_d;s)}\prod_{i=1}^dq_i(u_i)}{\sum_{u_1,\dots,u_d}e^{-yE_i(u_1,\dots,u_d;s)}\prod_{i=1}^dq_i(u_i)}\\
	U_a &= \sum_{h_1,\dots,h_k}\frac{E_a(h_1,\dots,h_k)e^{-yE_a(h_1,\dots,h_k)}\prod_{a=1}^kp_a(h_a)}{\sum_{h_1,\dots,h_k}e^{-yE_a(h_1,\dots,h_k)}\prod_{a=1}^kp_a(h_a)}\\
	U_{ia} &=\sum_{h,u}\frac{E_{ia}(h,u)e^{yE_{ia}(h,u)}p(h)q(u)}{e^{yE_{ia}(h,u)}p(h)q(u)}
\end{aligned}
\end{align}
Note that at finite $J$, the internal energy contains a term minimizing the number of violated constraints and a term minimizing the distortion between the source and the configuration (see equation \ref{eq:MinNrjcutoffJ}). 
We can expect that for $J$ large enough the cost of violating a clause (cost $2J$) becomes large compared to the distortion cost in such a way that the support of the measure $\mu$ is concentrated on configurations satisfying the constraints.
Finally, one can compute the zero-temperature complexity $\Sigma_e(y)$ from the 1RSB potential $\Phi_e^{\rm 1rsb}$ and the internal energy $U_{\rm int}^{\rm 1rsb}$:
\begin{align}
	\label{eq:complexity_energetic}
	\Sigma_e(y) =  \Phi_e^{\rm 1rsb}(y) + y U_{\rm int}^{\rm 1rsb}(y)
\end{align}

%% file: sections/sparsebasis.tex
\section{Sparseness of the basis}\label{sect:sparsebasis}

In this appendix we focus on the case where variable nodes have degree at most $2$. 
We build a basis spanning the set of codewords $\mathcal{C} = \{\underline{x} : H\underline{x}=0, \underline{x} \in \{0,1\}^{n}\}$ using a version of the Leaf Removal algorithm in which one chooses the variables to be removed according to their depth. 
We show that the Hamming weight of basis vectors is upper-bounded by the solution of a message-passing equation introduced in \cite{MontanariSemerjian06} by A. Montanari and G. Semerjian. 
Using the Replica Symmetric formalism, we can predict the value of this upper bound for random XORSAT instances with a given degree profile, in the thermodynamic limit.
When the RS prediction for the upper bound is finite, it indicates that the basis vectors constructed with this procedure have a sub-linear Hamming weight.
Therefore the basis is sparse, which means that the set of codewords is well connected, because from any codeword it is possible to reach another codeword by flipping a sub-linear number of variables (see the discussion in the introduction \ref{subsec:clustering_and_sparse_basis}).

\subsection{Construction of the basis}

The construction of the basis uses the Leaf Removal (LR) algorithm applied to the bipartite factor graph $\mathcal{G}=(V,F,E)$ with variable nodes in $V$, factor nodes in $F$, and edge set $E$. Variable nodes can have degree at most $2$. If all variables nodes have degree $2$, one removes a factor node (a row in the parity-check matrix $H$) in order to uncover some leaves. We will see at the end of this section (see \ref{subsec:1_reduction}) that it leaves the set of codewords $\mathcal{C}$ unchanged.
At each time step $t$, the LR algorithm removes a variable node $v_t$ among the leaves present in the graph, according to its depth. More precisely, the algorithm picks a variable uniformly at random among the leaves with smallest depth (i.e. closest to a leaf in the original graph). The unique factor node $a_t$ that was attached to $v_t$ is also removed, thus uncovering new leaves in the new factor graph.
Since in our setting variables have at most degree $2$, and assuming the graph to be connected, the core of $\mathcal{G}$ must be empty, therefore the procedure ends at $t=m$, when all the factor nodes in $F$ have been removed. The variables have been split in two sets: the $m$ variables removed $v_1, \dots, v_m$ are called the dependent ones, and the remaining ones are independent variables (denoted $w_1,\dots,w_{n-m}$).
This procedure allows to re-write $H$ in upper triangular form by a permutation of the rows and columns of $H$. The first $m$ columns correspond to the dependent variables $v_1,\dots,v_m$. The rows are ordered correspondingly: the $t^{\rm th}$ row corresponds to factor node $a_t$ (i.e $a_1,\dots,a_m$ from top to bottom). The last $n-m$ columns correspond to the $n-m$ independent variables (in arbitrary order).
This permutation transforms $H$ into a matrix $(T|U)$ with $T$ a $m\times m$ matrix in upper triangular form, and $U$ is a $m\times(n-m)$ matrix. 
Finally, one performs Gaussian elimination to transform the matrix $(T|U)$ into $(I_m|U')$, with $I_m$ the identity matrix of size $m$, and $U'=T^{-1}U$ a $m\times (n-m)$ matrix. The basis vectors of the codebook $\mathcal{C}$ are obtained as the $n-m$ columns of the following matrix of size $n\times(n-m)$, built with two blocks: on top there is $-U'$, below there is the identity $I_{n-m}$ of size $n-m$. 
Let $w$ be an independent variable, with $\partial w=\{a,b\}$ its two neighbors factor nodes. We call $p^*_{w\to a}$ (resp. $p^*_{w\to b}$) the shortest path going from $w$ to a leaf of $\mathcal{G}$, with the constraint that the first step goes through $a$ (resp. $b$).
We show in the following that the non-zero entries of the basis vector associated of $w$ correspond to the variables in the set $\{w\}\cup ( \{p^*_{w\to a}\cup  \{p^*_{w\to b}\}\setminus\{p^*_{w\to a}\cap p^*_{w\to b}\} )$. 

\subsection{Row operations}

During Gaussian elimination, one performs additions of the rows $L_1,\dots,L_m$ of the matrix $T$ to obtain the rows $L'_1,\dots,L'_m$ of the identity matrix $I_m$. By definition the row $L_m$ at the bottom does not have to be modified, it is already in the right form, thus $L'_m=L_m$. Looking at the row above $L_{m-1}$, if the unique element at the right of the diagonal term is non-zero then one performs the row addition $L_{m-1}' = L_{m-1}+L_m$. If not then it means that $L_{m-1}$ is already in the right form and there is no need to change it: $L_{m-1}' = L_{m-1}$. We can proceed in that way going from bottom row to top row. When one looks at the $i^{\rm th}$ row $L_i$, all the rows $L_j'$, with $j>i$ below are already in the right form (only the diagonal term is non zero). One has to look for the non-zeros elements of $L_i$, say they are at position $u>i$, $v>i$, $w>i$, then one needs to do the operation $L_i' = L_i+L_u'+L_v'+L_w'$.

\begin{figure}[htb]
	\begin{center}
		\includegraphics[width=5cm]{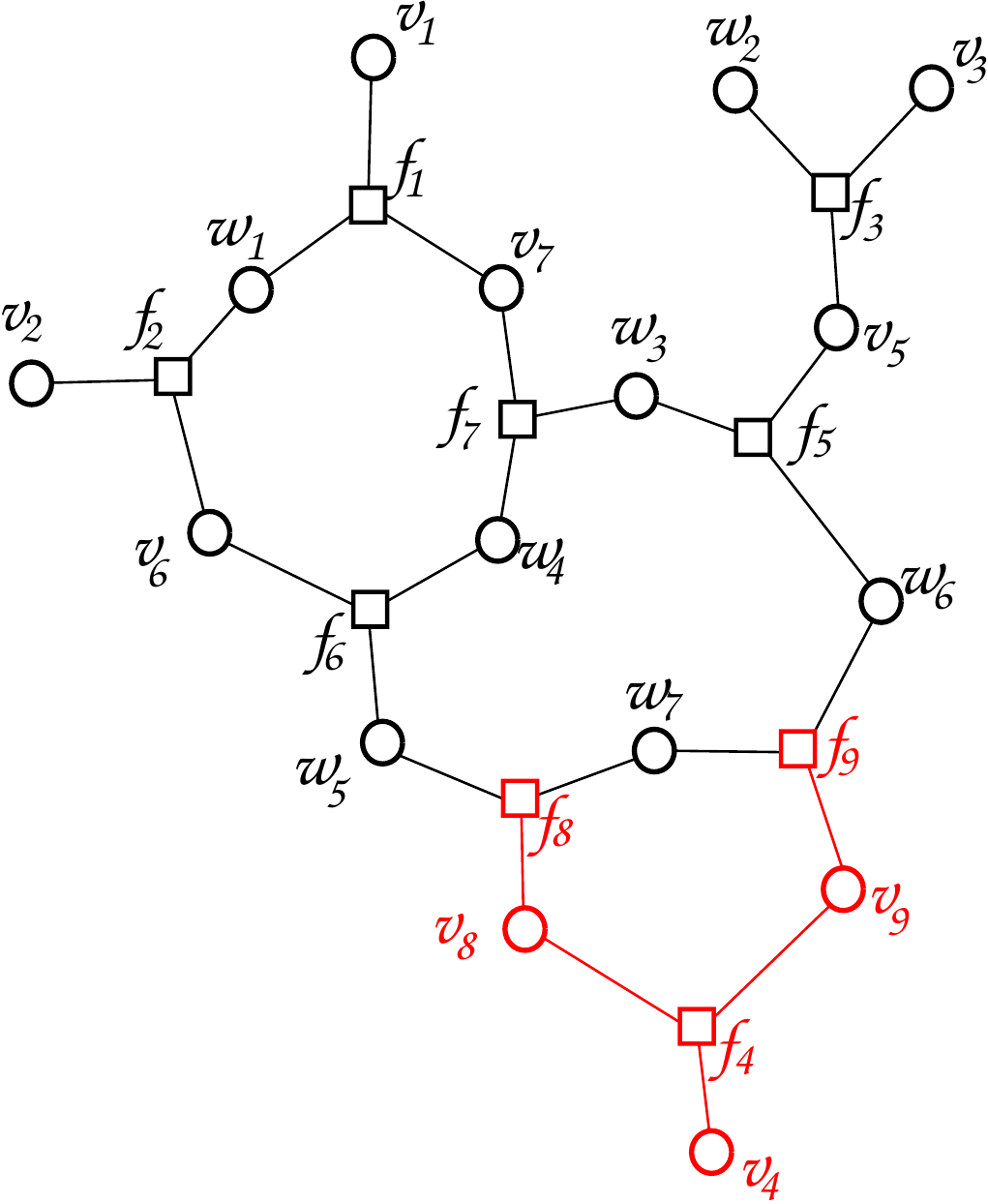}
		\caption{On a small example, the tree $B_4$ (in red) started from the variable $v_4$ removed at $t=4$. The row operation for the $4^{\rm th}$ row is thus $L_4'=L_4 + L_8 + L_9$}
		\label{fig:tree_B4}
	\end{center}
\end{figure}

The operation on each row can be written as $L_t'=L_t+\sum_{s\in B_t}L_s$, with $B_t$ a subset of $\{t+1,\dots,m\}$, that can be interpreted in terms of the paths explored by Leaf Removal. 
Say that at time $t-1$ during LR algorithm, the graph contains $q$ leaves with smallest depth $d$, denoted $\{u_1,\dots,u_q\}$, and possibly other leaves with larger depths. 
At time $t$, the algorithm picks one leaf with smallest depth $d$: $v_t=u_i$ for some $i\in\{1\dots,q\}$. 
The variable $v_t$ is removed, as well as the unique factor node $a_t$ attached to $v_t$ at time $t-1$. 
The degree of the variables in $\partial a_t\setminus v_t$ is decreased by $1$.
Those that had degree $1$ at time $t-1$ become of degree $0$ at time $t$, and are added to the set of independent variables. 
Those that had degree $2$ at time $t-1$ become leaves at time $t$, and can be removed by the algorithm at later times.
These variables have depth exactly $d+1$. 
Indeed they cannot have a degree $>d+1$ because they are connected to $v_t$ which is of degree $d$.
They cannot have a degree $<d$, otherwise they would have been removed at a previous time.
They cannot have degree exactly $d$, because in that case they would be connected (in the original graph) to another variable of depth $d-1$, which therefore has been removed at a previous time $<t$ and this would mean that they have degree $1$ at time $t-1$.
The set $B_t$ can be defined recursively as the set of all the variable nodes that have been visited by a path starting from node $v_t$. 
More precisely, $B_t$ is a tree of root $v_t$, and the variables in the first generation of $B_t$ are the variables $\partial a_t\setminus v_t$ that have depth $d+1$ and that has been picked by the algorithm at times $t'>t$. 
A variable $v_{t"}$ of the second generation have depth $d+2$, and is linked to a variable $v_{t'}\in \partial a\setminus v_t$ that belongs to the first generation. The variable $v_{t"}$ has become a leaf at time $t'$, and have then been visited by the algorithm at time $t">t'$. 
The next generations are built similarly, the variables of the $r^{\rm th}$ generation in $B_t$ having depth $d+r$. 
The tree ends in dependent variables $v$ that are surrounded by variables that have not been picked by the algorithm (i.e. that belongs to the set of independent variables $w_1,\dots,w_{n-m}$). The tree $B_t$ is represented in the figure \ref{fig:tree_B4} on a small example.
If at a later time $t<s\leq t+q$ another leaf $u_j$ in $\{u_1,\dots,u_q\}$ is explored: $v_s=u_j$, then a new subset $B_s$, disjoint from $B_t$, will be built starting from $v_s$.

\subsection{Effect of row operations on $U$}

We want to know how many non-zero entries are created in $U'$ when the row operations $L'_{t}=L_t+\sum_{s\in B_t}L_s$ are performed. The columns of $U$ correspond to the independent variables $w_1,\dots,w_{n-m}$ that have not been removed by the LR algorithm. This means that at some time step the degree of these variables has become zero. One can distinguish between two cases: either the independent variable $w_i$ had degree $2$ in the initial graph $\mathcal{G}$, or it had degree $1$.

\begin{figure}[htb]
	\begin{center}
		\includegraphics[width=5cm]{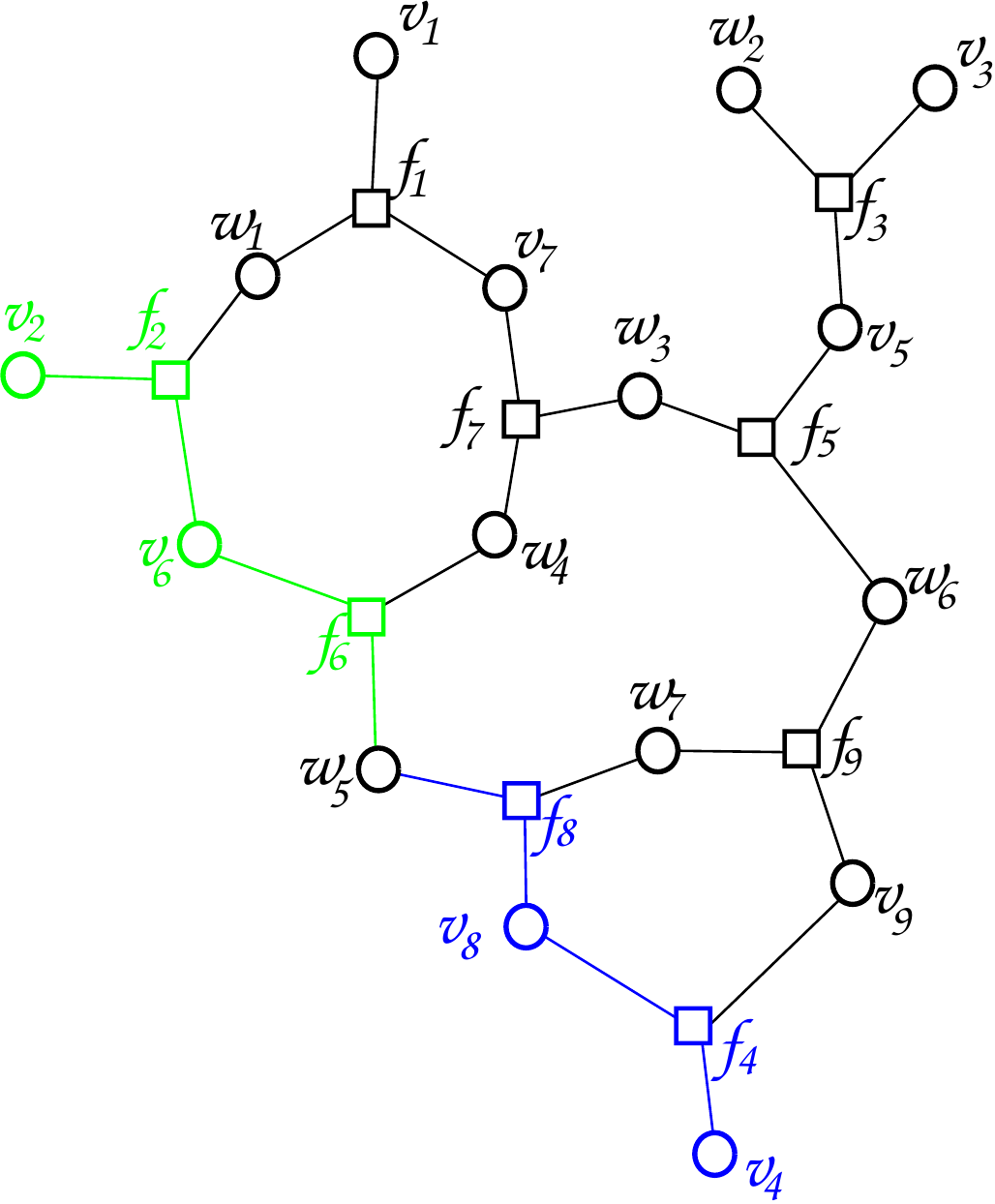}
		\caption{The two paths $\mathcal{P}_{2\to 6}=\{v_2, v_6\}$ (in green) and $\mathcal{P}_{4\to 8}=\{v_4, v_8\}$ (in blue) going from leaves $v_2$ and $v_4$ toward the independent variable $w_5$. The non-zero entries in the $5^{\rm th}$ column of $U'$ corresponding to $w_5$ are at the $2, 4, 6$ and $8^{\rm th}$ positions.}
		\label{fig:paths}
	\end{center}
\end{figure}

\begin{enumerate}
	\item In the first case let $s,t$ be the two time steps at which the two factor nodes $a_s$ and $a_t$ linked to $w_i$ have been removed. In the $i^{\rm th}$ column of $U$ there is $2$ non-zero entries at positions $s$ and $t$. The variable $v_s$ (resp. $v_t$) that have been removed at $s$ (resp. at $t$) belongs to $\partial a_s\setminus w_i$ (resp. to $\partial a_t\setminus w_i$). Moreover, there is an unique $s_0<s$ (resp. an unique $t_0<t$) such that $v_{s_0}$ (resp $v_{t_0}$) is a leaf in the initial graph, and such that $v_s\in B_{s_0}$ (resp. $v_t\in B_{t_0}$). It might be that these two leaves coincides: $v_{s_0}=v_{t_0}$. Let $\mathcal{P}_{s_0\to s}\subseteq B_{s_0}$ (resp. $\mathcal{P}_{t_0\to t}\subseteq B_{t_0}$) be the path going from $v_{s_0}$ to $v_s$ in the tree $B_{s_0}$ (resp. from $v_{t_0}$ to $v_t$ in the tree $B_{t_0}$). 
	\begin{enumerate}
		\item if $v_{s_0}\neq v_{t_0}$ then one can check that in the $i^{\rm th}$ column of $U'$ there is non-zero entries in all the positions corresponding to the variables belonging to one of the two paths $\mathcal{P}_{s_0\to s}$ and $\mathcal{P}_{t_0\to t}$. This is the case for the variable $w_5$ (see figure \ref{fig:paths}).
		\item if $v_{s_0}= v_{t_0}$ then the beginning of the two paths coincides for some variables $\{v_{s_0}\to \dots\to v_r\}=\mathcal{P}_{s_0\to s}\cap\mathcal{P}_{t_0\to t}$. After some factor node $a_r$ the paths splits in two parts. 
		The subset $\{w_i\}\cup\mathcal{L}_i$ with $\mathcal{L}_i=(\mathcal{P}_{s_0\to s}\cup\mathcal{P}_{t_0\to t})\setminus(\mathcal{P}_{s_0\to s}\cap\mathcal{P}_{t_0\to t})$ forms a loop (or a set of disjoint loops) that passes through $a_s, a_t$ and $a_r$. One can check that the non-zero entries in the $i^{\rm th}$ column of $U'$ coincides with the variables belonging to $\mathcal{L}_i$. This is the case for the variable $w_7$ (see figure \ref{fig:paths}, with $\mathcal{L}_7=\{v_8,v_9\}$)
	\end{enumerate}
	\item In the second case $w_i$ has degree $1$ in the initial graph (i.e $w_i$ has depth $0$). Let $t$ be the time at which the unique factor node $a_t$ linked to $w_i$ has been removed. Since the LR algorithm picks always variables with smallest depth, this means that the variable $v_t$ that has been picked by the algorithm has also depth $0$.). Therefore in this case the path is reduced to the unique variable $v_t$: $\mathcal{P}_{t\to t}=\{v_t\}$. In the $i^{\rm th}$ column of $U'$, there is a unique non-zero entry at the $t^{\rm th}$ position, and zeroes elsewhere. This is the case for the variable $w_2$ on figure \ref{fig:paths}, the associated column in $U'$ has a non-zero entry at position $t=3$. 
\end{enumerate}
By construction, $\cP_{s_0\to s}$ (resp. $\cP_{t_0\to t}$) is a path of strictly decreasing depth, going from $w_i$ to a leaf $s_0$ (resp. $t_0$), where the first step goes through $a_s$ (resp. $a_t$).
Since a path of strictly decreasing depth must be the shortest path to a leaf, $\cP_{s_0\to s}$ is actually the shortest path from $w$ to a leaf, with the first step going through $a_t$: $\cP_{s_0\to s}=p^*_{w_i\to a_s}$ using the notations introduced in the first section of this appendix (and similarly $\cP_{t_0\to t}=p^*_{w_i\to a_t}$).
Note that we restrict ourselves to non-backtracking paths, i.e. paths of the form $p=(p(1)=i_1, p(2)=a_1, \dots, p(2w-1))=i_w$ (alternating variable nodes and factor nodes), such that $p(t)\neq p(t+2)$ $\forall t$.

\subsection{Upper bound}

We just showed that the basis vector $b_w$ relative to the independent variable $w$ is of the form $b_w = \underline{x}_{p^*_{w\to a}} \oplus \underline{x}_{p^*_{w\to b}} \oplus \underline{x}_{\{w\}}$ where $\partial w = \{a,b\}$, and $p^*_{w\to a}$ (resp. $p^*_{w\to b}$) is the shortest path going from $w$ to a leaf, where the first step goes through $a$ (resp. $b$).
We have used the notation $\underline{x}_I$, with $I\subseteq V$, where the vector $\underline{x}_I$ contains non-zero entries only at indices $i\in I$.
The Hamming weight of the basis vector $b_w$ relative to independent variable $w$ can be upper bounded as follows: 
$$w(b_w) \le w(\underline{x}_{p^*_{w\to a}}) +  w(\underline{x}_{p^*_{w\to b}}) \quad .$$
Therefore, if the two paths $p^*_{w\to a}, p^*_{w\to b}$ have length $o(n)$, then so do the basis vector, making the basis sparse. 
Note that when measuring the length of a path or a distance, we count $+1$ every time a variable node is met (i.e. we do not count the factor nodes present in the path). 

The length $w(p^*_{i\to a})$ for any variable node $i\in V$ can actually be computed exactly as the solution of the message-passing equations for the minimal size rearrangement, defined in equation (36), section V.A. of \cite{MontanariSemerjian06}, that we re-write here with an additional update-scheme:
\begin{equation}
	\begin{aligned}
		h_{i\to a}^{(t+1)} &= 1 + \sum_{b\in\partial i \setminus a}\min_{j\in\partial b \setminus i} h_{j\to b}^{(t)} \\
		h_i^{(t)} &= 1 +  \sum_{b\in\partial i} \min_{j\in\partial b \setminus i} h_{j\to b}^{(t)}
	\end{aligned}\label{eq:mspath}
\end{equation}
where the set $b\in\partial i \setminus a$ is made of either one or zero elements, because the degree of variables is $\le 2$. 
As usual, we use the convention that summing over an empty set gives zero, and the minimum over an empty set is $+\infty$.
In case of a tree, the quantity $h_i$ corresponds exactly to the weight of the minimal size rearrangement, which is the smallest set of variables that are needed to be flipped when the variable $x_i$ is flipped.
Notice that these equations are nothing but the Max-Sum updates for our problem \eqref{eq:MS} with all $s_i=+1$, i.e. with the zero codeword as the source.
We initialize all messages to $h_{i\to a}^{(0)} = u_{a\to i}^{(0)} = +\infty$.

\begin{figure}
     \centering
     \includegraphics[width=\columnwidth]{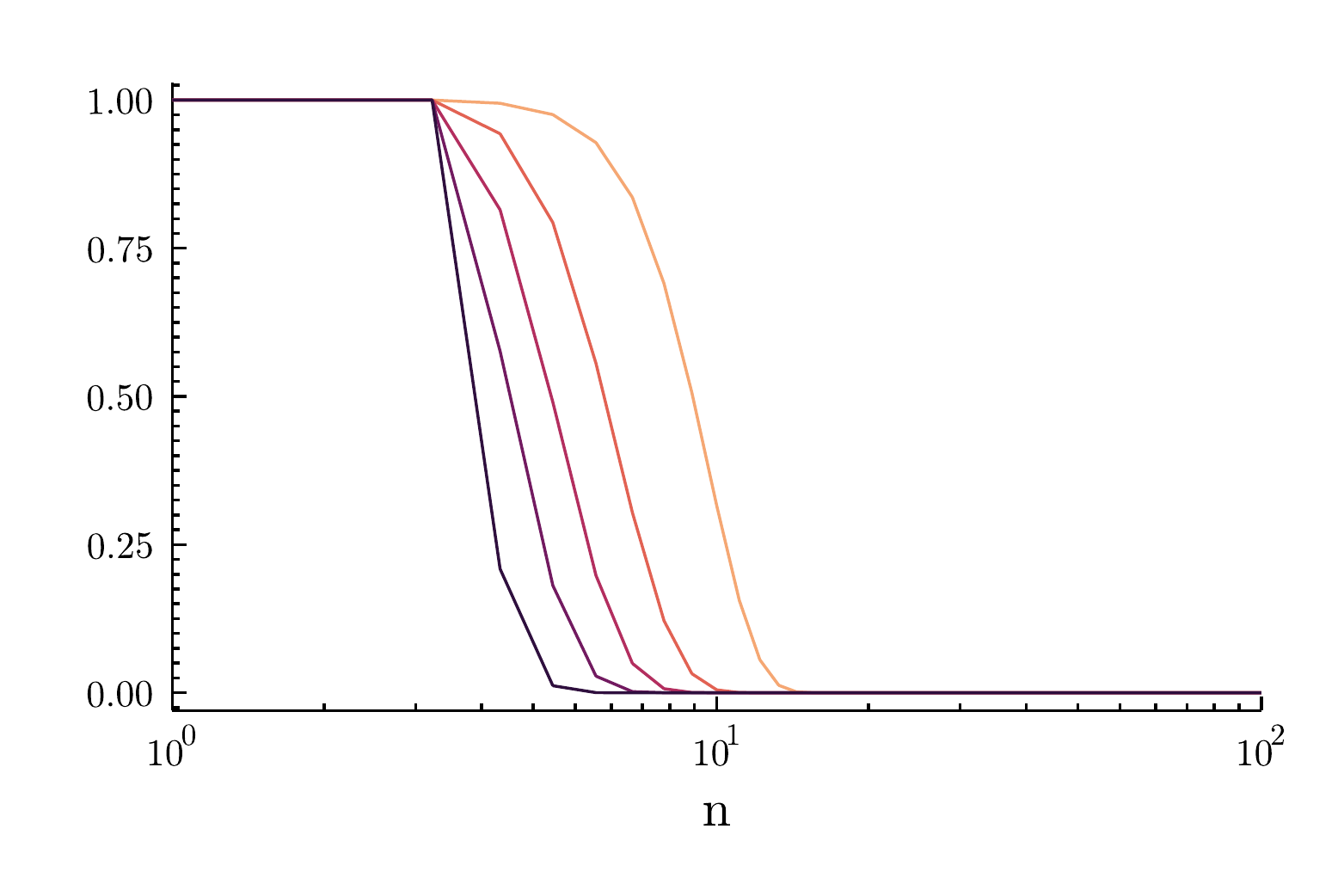}
     \caption{Cumulative probability distribution $p(n)=P(\text{Rearrangement size} \ge n)$ obtained via the RS computation for fixed factor degree $k=3$. Curves correspond to increasing values of $\lambda_1=\{0.1,0.3,0.5,0.7,0.9\}$ from right to left.}
     \label{fig:rearrangement_hist}
 \end{figure}

In the following we will prove that $h_{i\to a}^{(t)}$ converge on any graph (not only on trees) to $1$ if $i$ has degree $1$, and to $w(p^*_{i\to b})$ if $\partial i = \{a, b\}$.
Iteration of \eqref{eq:mspath} resembles the breadth-first search algorithm, with the slight complication that it applies to non-backtracking paths. 
Consider an edge $(ia)$. Let $p_{(ia)}=(v_1=j,..., \ldots,v_{k-1}=i,v_k=a)$ be the shortest non-backtracking path (i.e. such that $v_t\neq v_{t+2}~\forall t$) starting from a (variable) leaf and ending in the two vertices $i,a$ (in this order). 
If $i$ is of degree $1$, then $p_{(ia)}=\{i,a\}$ and $w(ia)=1$. If instead $i$ is of degree $2$, then let $b$ be the other factor node attached to $i$, in that case the weight of the path $p_{(ia)}$ equals the weight of the path $p^*_{i\to b}$: $w(ia)=w(p^*_{i\to b})$.
Define $w(ia)=\frac{k}2$ the number of variable vertices in this path. 
It is easy to see that $h_{i\to a}^t$ is non-increasing in $t$, as the expression is monotonous and can only decrease in the first step because the starting point is all $+\infty$.
We show now that (a) $h^t_{i\to a}=w(ia)$ for $t\geq w(ia)$ and (b) $h^t_{i\to a}=+\infty$ for $t<w(ia)$. Let us proceed by induction in $t$. The result is clearly true for $t=1$, because $w(ia)=1$ implies that $i$ is a leaf, but then the r.h.s. of \eqref{eq:mspath} is $1$. 
Suppose the statement true for every $t'<t$ and take $(ia)$ with $w(ia) = t$. Consider its corresponding shortest non-backtracking path $p=(...,j,b,i,a)$ starting on a leaf an ending in $(i,a)$ containing $w(ai)$ variable vertices. Then:
\begin{itemize}
	\item $w(jb)=w(ai)-1$ because the sub-path ending in $j,b$ must be optimal, so $h_{j\to b}^{(t-1)} = w(jb)$ by IH,  and
	\item $w(kc)\geq w(ai)-1$  for any $k\in \partial c\setminus i$, $c\in \partial i\setminus a$ as otherwise, we could construct a strictly shorter path from a leaf to $i,a$. As $h_{k\to c}^{(t-1)}$ is either $w(kc)$ or $+\infty$ by IH, this implies $h_{k\to c}^{(t-1)}\geq w(ai)-1$.
\end{itemize}
By means of \eqref{eq:mspath}, we conclude $h^t_{i\to a} = w(ia)$. Similarly take $(ia)$ with $t<w(ia)$; we know that $w(kc)\geq w(ai)-1 > t-1$ for all $k\in \partial c\setminus i$, $c\in \partial i\setminus a$ so by induction hypothesis the RHS of \eqref{eq:mspath} is $+\infty$ and then $w(ia)=+\infty$. 
As a consequence, \eqref{eq:mspath} converges in exactly $\max_{(ia)\in E} w(ia) \leq |E|$ steps.
Applying the RS version of \eqref{eq:mspath} for the ensemble of cycle codes with factor degree fixed to 3 and a fraction $\lambda_1$ of degree-1 variables gives the result in Fig.\ref{fig:rearrangement_hist}.
It is clear that all of the probability mass is found at finite rearrangement size.
Since rearrangements are an upper bound for the Hamming weights of basis vectors, we conclude that the latter must be finite.

\subsection{Effect of $1$-reduction}
\label{subsec:1_reduction}

If all variables in $\mathcal{G}$  have even degrees (as is the case with all variables of degree 2), then the associated codebook is left unchanged when one removes only one factor node $a$ (i.e. when a $1$-reduction is performed). The reason is simple: the sum in $\text{GF}(2)$ of all rows in the matrix $H$ gives the all-0 row, implying that any given row in H can be built as the sum of the other rows and thus is linearly dependent on them.

%% file: sections/ldpc.tex
\section{Relation to LDPC decoding}\label{sect:ldpc}

\begin{figure}[t]
	\begin{center}
		\includegraphics[width=0.7\columnwidth]{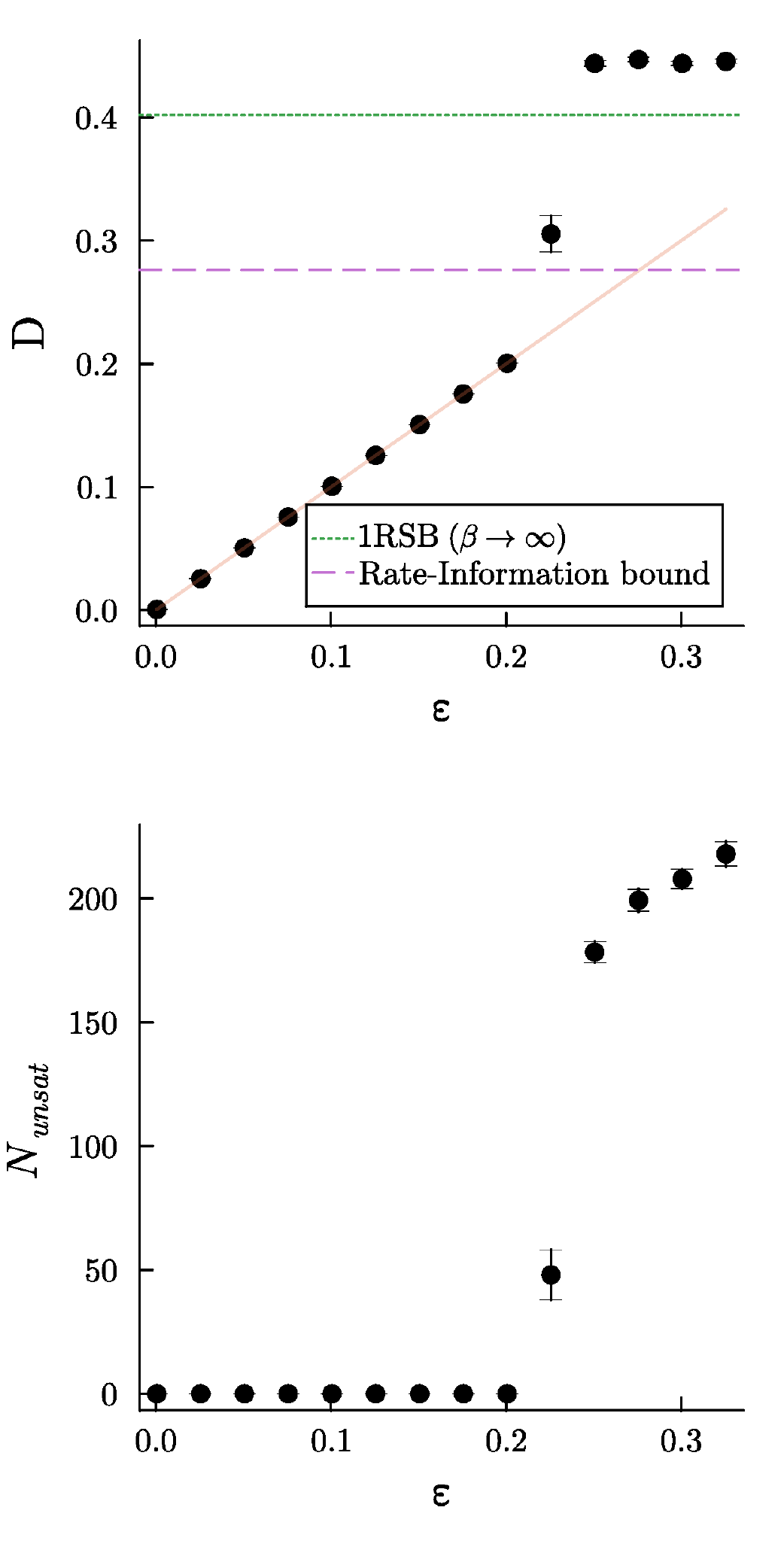}
		\caption{Performance of Max-Sum with reinforcement on source vectors made by perturbing the all-zeros codeword with $\varepsilon n$ bit flips. $n=1800, \Lambda=\{\lambda_2=0.45, \lambda_3=0.55\}, P=\{p_3=1\}, R=0.15$, average over 50 instances.
        The upper panel shows the average distortion, with the Shannon bound and 1RSB prediction for a symmetric source as references. The lower panel shows the number of unsatisfied factors.}
		\label{fig:planted}
	\end{center}
\end{figure}

The problem of channel coding is formally identical to source coding in the sense that it is stated as the minimization of the same energy function \eqref{eq:MinNrjcutoffJ} (see e.g. \cite[equation 3]{nakamura_statistical_2001}.
However, there are two main differences that make one problem substantially harder than the other. 
The first difference is a design choice: a set of codewords which is good for channel coding is typically awful for source coding. 
In the communication problem one wants codewords to be as far as possible from one another, while for compression one wants them covering the whole space, so that any point is close enough to at least one codeword.
The second difference lies in the distribution of the source vector.
While here we are taking every bit of the source uniformly at random, in channel coding the source vector is made of a codeword with a small fraction of corrupted bits.
This means that, provided the code is well-designed, there will always be one codeword significantly closer to the source than others, making the problem typically easier.

The results of the two problems are not directly comparable, however we give in Fig. \ref{fig:planted} some evidence that the compression problem becomes easy when the source vector is close enough to a codeword.
We place ourselves in a regime where the compression problem with symmetric source is in a 1RSB phase, but modify the source starting from a codeword and then flipping a fraction $\varepsilon$ of the bits.
The compression is then performed by Max-Sum with reinforcement.
For small values of $\varepsilon$, Max-Sum is able to solve the problem exactly, as the number of unsatisfied factors is zero and the distortion is equal to $\varepsilon$.
Indeed, by construction, $\varepsilon$ is the (normalized) Hamming distance from the starting codeword. By definition, the closest codeword will always be at a distance smaller or equal than $\varepsilon$. As $\varepsilon$ increases, we observe a transition where the algorithm stops converging to solutions, and the vectors found (after fixing a set of independent variables as described in section \ref{subsec:deg2_cavityresults}) have a distortion which is even worse than the 1RSB prediction for the same graph (for a symmetric source).